\documentclass[twocolumn,prd, aps, notitlepage, nofootinbib, showpacs, superscriptaddress,longbibliography]{revtex4-1}
\usepackage[utf8]{inputenc}
\usepackage{graphicx,mathtools,xcolor,latexsym,rotating,soul}
\usepackage[T1]{fontenc}
\usepackage{tikz}
\usepackage{comment}
\usepackage{amsmath, amssymb, amsthm, amsfonts,breqn}
\usepackage{multirow}
\usetikzlibrary{decorations.pathmorphing}
\usepackage[colorlinks=true,citecolor=green,urlcolor=blue,linkcolor=blue]{hyperref}
\usepackage{pifont}
\usepackage{gensymb}
\usepackage{enumitem}
\usepackage{mathrsfs}
\usepackage{amsbsy}
\usepackage[scr=boondox,  
            cal=esstix]   
           {mathalpha}

\newtheoremstyle{named}{}{}{\itshape}{}{\bfseries}{.}{.5em}{\thmnote{#3 }#1}
\theoremstyle{named}

\newcommand*{\rom}[1]{\expandafter\@slowromancap\romannumeral #1@}

\newcommand{\DI}{\text{DI}}

\makeatletter
\let\cat@comma@active\@empty
\makeatother
\allowdisplaybreaks
\begin{document}
\title{Relativistic mode sums for neutron-star tidal response}
\author{Abhishek Hegade K. R.}
\email{ah4278@princeton.edu}
\affiliation{Princeton Gravity Initiative, Princeton University, Princeton, NJ 08544, USA}
\author{K.J. Kwon }
\email{jameskwon@ucsb.edu}
\affiliation{Department of Physics, University of California, Santa Barbara, CA 93106, USA}
\author{Tejaswi Venumadhav }
\email{teja@ucsb.edu}
\affiliation{Department of Physics, University of California, Santa Barbara, CA 93106, USA}
\affiliation{International Centre for Theoretical Sciences, Tata Institute of Fundamental Research, Bangalore 560089, India}
\author{Hang Yu}
\email{hang.yu2@montana.edu}
\affiliation{eXtreme Gravity Institute, Department of Physics, Montana State University, Bozeman, MT 59717, USA}
\author{Nicolas Yunes}
\email{nyunes@illinois.edu}
\affiliation{Illinois Center for Advanced Studies of the Universe, Department of Physics, University of Illinois at Urbana-Champaign, Urbana, IL 61801, USA}
\begin{abstract}
Time-dependent tidal interactions during the late inspiral of binary neutron stars encode valuable information about neutron-star structure, but systematically extending the familiar Newtonian mode-sum picture into full general relativity is nontrivial. In this paper, we develop a practical relativistic implementation of mode-sum tidal response for non-rotating neutron stars in Regge-Wheeler gauge. Using near-zone boundary conditions, we systematically define the interior tidal field, the relativistic overlap integrals, and the corresponding mode amplitudes. The good is that the dominant f-mode contribution is remarkably robust, reproducing the direct matching calculation to within $\sim 3$\% across the equations of state we consider. The bad is that the operator governing mode inner product is not positive definite on the full Regge-Wheeler-gauge function space, so the relativistic mode sum truncated at ${\cal{O}}(\omega^2)$ is not expected to strictly converge to the direct matching solution. The subtle is that the tidal field inside the star is not unique, although this ambiguity has only a limited impact on the dominant f-mode response for the classes of extensions studied here. Our results establish the practical utility of relativistic mode-sum approximations, while making clear that their predictive power comes from a controlled low-mode description, rather than from a formally convergent strong-field expansion.
\end{abstract}
\maketitle
\section{Introduction}
A neutron star in a binary system is tidally deformed by its companion. The tidal deformations change the observed gravitational waves emitted by the binary, and can help constrain the equation of state (EoS) of a neutron star~\cite{Flanagan:2007ix,LIGOScientific:2018cki,Chatziioannou_2020}.
During the early phase of the inspiral, the tidal deformations are approximately quasi-static, and the tidal response of the neutron star is quantified by a single number, called the \textit{static tidal deformability}~\cite{Flanagan:2007ix}.
A well-established formalism for modeling the static tidal deformability exists in general relativity~\cite{Hinderer:2007mb,Binnington:2009bb,Damour:2009vw}, allowing the exploration of how different nuclear physics properties affect the time-independent tidal deformation of a neutron star. Extending this approach to frequency-dependent tidal deformations in general relativity is currently an active area of research~\cite{Chakraborty:2026qru}.

In Newtonian gravity, the dynamical tidal response of a non-rotating star to an external tidal field can be modeled as a mode sum, with each stellar mode treated as an ``effective’’ harmonic oscillator. This simplification is possible due to the self-adjoint nature of the linear perturbations~\cite{1964ApJ...139..664C,1967MNRAS.136..293L,1978ApJ...221..937F} and the ease with which one can separate  the tidal field inside the star from the self-field~\cite{Press-Teukolsky}.
The presence of a complete set of eigenfunctions in Newtonian theory has resulted in a large body of literature, where linear and non-linear tidal interactions are analyzed by modeling linear and coupled non-linear oscillator problems~\cite{Schenk_2001,Weinberg_2012,Yu_2022,Kwon:2024zyg,Kwon:2025zbc}.

Motivated by Newtonian theory, several studies~\cite{Hinderer:2016eia,Steinhoff:2016rfi, Steinhoff:21, Yu:25a, Haberland:25} have used a phenomenological approach to model the frequency-dependent tidal response function that adopts a Newtonian-like mode approximation in effective one-body models, which are calibrated to numerical relativity simulations. 
While this approach has been successfully applied in waveform modeling~\cite{Abac_2024}, its phenomenological nature obscures the errors associated with the mode-sum approximation and its calibration. In addition to this phenomenological approach, there have also been several studies that use techniques from scattering amplitudes and black hole perturbation theory to model the time-dependent response using the quasi-normal modes (QNM) of a neutron star~\cite{chakrabarti2013newperspectivesneutronstar,Saketh:2024juq,Miao:2025utd}.

Alternatively, one can use tools from matched-asymptotic expansions and post-Newtonian (PN) theory~\cite{Poisson:2020vap} to model the frequency-dependent tidal response function of a neutron star. 
A PN approximation is one in which the Einstein equations and their approximate solutions are expanded in weak fields and small velocities~\cite{Blanchet:2013haa}, while matched-asymptotic expansions are an applied-mathematics technique to construct global solutions to partial differential equations by combining approximations valid in different, but overlapping regimes~\cite{Bender:1999box}. In this approach, one can understand how the PN dynamics of the orbit affects the tidal deformations of the binary stars, and systematically improve the approximation by including higher PN-order corrections. Recently, several groups have adopted this approach to capture various novel features in the tidal response function present in general relativity, such as mode resonances~\cite{Pitre:2023xsr,HegadeKR:2024agt,Andersson:2025iyd}, tidal dissipation~\cite{Ripley:2023lsq,Ripley:2023qxo,HegadeKR:2024agt}, mode expansions~\cite{HegadeKR:2025qwj,Martinez-Rodriguez:2026omk}, nonlinear tidal interactions~\cite{Pitre:2025qdf,pani2025nonlinearrelativistictidalresponse}, quasi-universal relations~\cite{Saes:2025jvr}, and to include the effects of damping due to radiation~\cite{Andersson:2025iyd}.

The main technical step in~\cite{Pitre:2023xsr,HegadeKR:2024agt,Andersson:2025iyd} relies on using solutions to the vacuum Einstein equations outside the star as a boundary condition at the surface of the star.
In such a matching procedure, the tidal response is obtained (numerically) as a function of the driving frequency, and the stellar mode frequencies can be extracted by numerically scanning the tidal response function to search for the poles of the function.
The tidal response function obtained through this matching procedure is functionally similar to a mode sum. Such a similarity prompted~\cite{Andersson:2025iyd} to numerically fit a mode sum to the response function by calculating the residue of the tidal response function near the mode frequencies. 
The mode amplitudes in this approach, the so-called overlap integrals, were an output of the fitting procedure, rather than a fundamental prediction.

In this paper, we present a systematic approach to construct a mode-sum approximation to the tidal response function of a neutron star in general relativity. Our approach seeks to answer the following questions raised in~\cite{Pitre:2023xsr}:
\begin{enumerate}
    \item What eigenfunction basis underlies the mode sum?
    \item How do we define the tidal field inside the star?
    \item What is the definition of the overlap integral in full general relativity?
\end{enumerate}
We attempted to answer these questions in harmonic gauge in a
short Letter~\cite{HegadeKR:2025qwj}, which we summarize below:
\begin{enumerate}
    \item The relevant basis is the set of stellar perturbation eigenfunctions that satisfy near-zone PN boundary conditions in the buffer zone, rather than quasi-normal boundary conditions at null infinity. This basis can accurately capture the QNM spectrum of compact stars up to the $f$-mode frequency~\cite{Lindblom_1997,Andersson:2025iyd}.
    \item The tidal field inside the star is defined by separating the metric perturbation inside the star arbitrarily into a self-field generated by the star and a tidal field. Any tidal field that satisfies the Hamiltonian and momentum constraint and matches smoothly to the external tidal field is a valid definition for the tidal field inside the star.
    \item A prescription for the overlap integral follows from the operator equation that describes the evolution of the eigenfunctions, once a choice of interior tidal field has been made.
\end{enumerate}
This formalism, however, has not yet been backed up with concrete numerical results in practice. That is the purpose of this paper. 

In this paper, we implement the framework of~\cite{HegadeKR:2025qwj} numerically in Regge--Wheeler gauge and use it to ask a sharper question: what survives of the Newtonian oscillator picture in the strong-field regime?

Our formalism suggests that there are three different aspects to consider here. The good is that the near-zone construction captures the dominant $f$ mode accurately. We find that the corresponding eigenfunctions and eigenfrequencies closely track the real part of the QNM solutions up to the $f$-mode frequency, and that the resulting $f$-mode contribution reproduces the direct matching calculation at the few-percent level across the equations of state studied here. The bad is that, unlike in Newtonian theory or in outgoing-null gauge~\cite{FS_stability_rel}, the norm operator that governs mode orthogonality in the strong-field problem is \textit{not} positive definite on the full Regge--Wheeler-gauge function space. As a result, the relativistic mode sum with near-zone boundary conditions truncated at $O(\omega^2)$ is not expected to converge strictly to the direct matching calculation implemented in~\cite{HegadeKR:2024agt}. The subtle point is that the tidal field inside the star is not unique: one may extend the exterior tidal field into the stellar interior in more than one consistent way, and different choices lead to different overlap integrals and different rates of convergence, even though the dominant $f$-mode response remains comparatively robust for the classes of extensions considered here.

It is important to note that we have made several assumptions when \textit{attempting} to answer the questions asked above. Our description is tied to the Regge-Wheeler gauge and it relies on the use of near-zone, low-frequency approximation to the tidal field outside the star. These approximations are valid during the late inspiral of non-spinning neutron stars on quasi-circular orbits. The results of this paper show that the overlap integral formulation in general relativity is much more subtle than Newtonian gravity. In fact, the issues we raise here are not high-frequency artifacts.
They are already visible at 1PN level~\cite{1964ApJ...139..664C}, where one can see the non-positive nature of the norm.

In arriving at the conclusions described above, we also obtain a variety of useful intermediate results and establish several ancillary conclusions. The eigenfunctions and eigenfrequencies obtained with near-zone boundary conditions closely track the real part of the QNM solutions for mode frequencies below the $f$-mode frequency, which we show is consistent with~\cite{Lindblom_1997,HegadeKR:2024agt,Andersson:2025iyd}. The operator equation we use to establish mode orthogonality is derived from~\cite{FN-book} and is closely related to the Detweiler--Ipser operators~\cite{1973ApJ...185..685D}, although important differences arise because the domain has finite size. We derive the ordinary differential equations inside the star that can be integrated to construct the tidal field, and we use these ingredients to compute the dynamical tidal response function in a mode-sum approximation that retains modes up to the $f$ mode. We illustrate these points with toy models that capture the same qualitative features as the relativistic problem, including both the ambiguity in extending the tidal field into the interior and the obstruction to strict convergence of the mode sum.

The remainder of this paper fills in the details behind the conclusions summarized above and is organized as follows. In Sec.~\ref{sec:toy-models}, we introduce toy models that exemplify our techniques and their limitations, and that help clarify the origin of the relativistic results. Section~\ref{sec:response-theory} contains the technical details of the perturbation theory, the construction of the tidal field inside the star, and the relativistic overlap integrals. In Sec.~\ref{sec:results}, we present our results, compare the near-zone eigenfunctions to the QNM solutions, and analyze the difference between the direct matching calculation and the mode-sum approximation for different choices of the interior tidal field. Our conclusions are presented in Sec.~\ref{sec:conclusions}. We use geometric units $G=1=c$ throughout the paper, and the metric signature is $(-,+,+,+)$.

\section{Toy models}\label{sec:toy-models}

In this section, simple toy models are used to illustrate our approach to relativistic tidal excitations. The toy models will clarify several important issues we encounter when extending the Newtonian-like mode-sum approach to general relativity.
In Sec.~\ref {sec:sturm-liouville-problem}, we use a Sturm-Liouville problem with non-homogeneous boundary conditions to illustrate how different choices for extension of the non-homogeneous boundary condition can lead to differences in the rate of convergence of the eigenfunction expansion.
Building on this, we next analyze a frequency-dependent Sturm-Liouville problem in Sec. ~\ref{sec:frequency-dependent-sturm-liouville-problem} to show how one can use a small frequency expansion to obtain a set of complete eigenfunctions if the operators are positive-definite.
Finally, we also use this example to show that when the operators are not positive-definite, as in general relativity, the mode-sum approximation does not converge.
\subsection{Sturm-Liouville problem with non-homogeneous boundary conditions and eigenfunction expansion}\label{sec:sturm-liouville-problem}

\begin{figure}[t]
    \centering
    \includegraphics[width=0.99\linewidth]{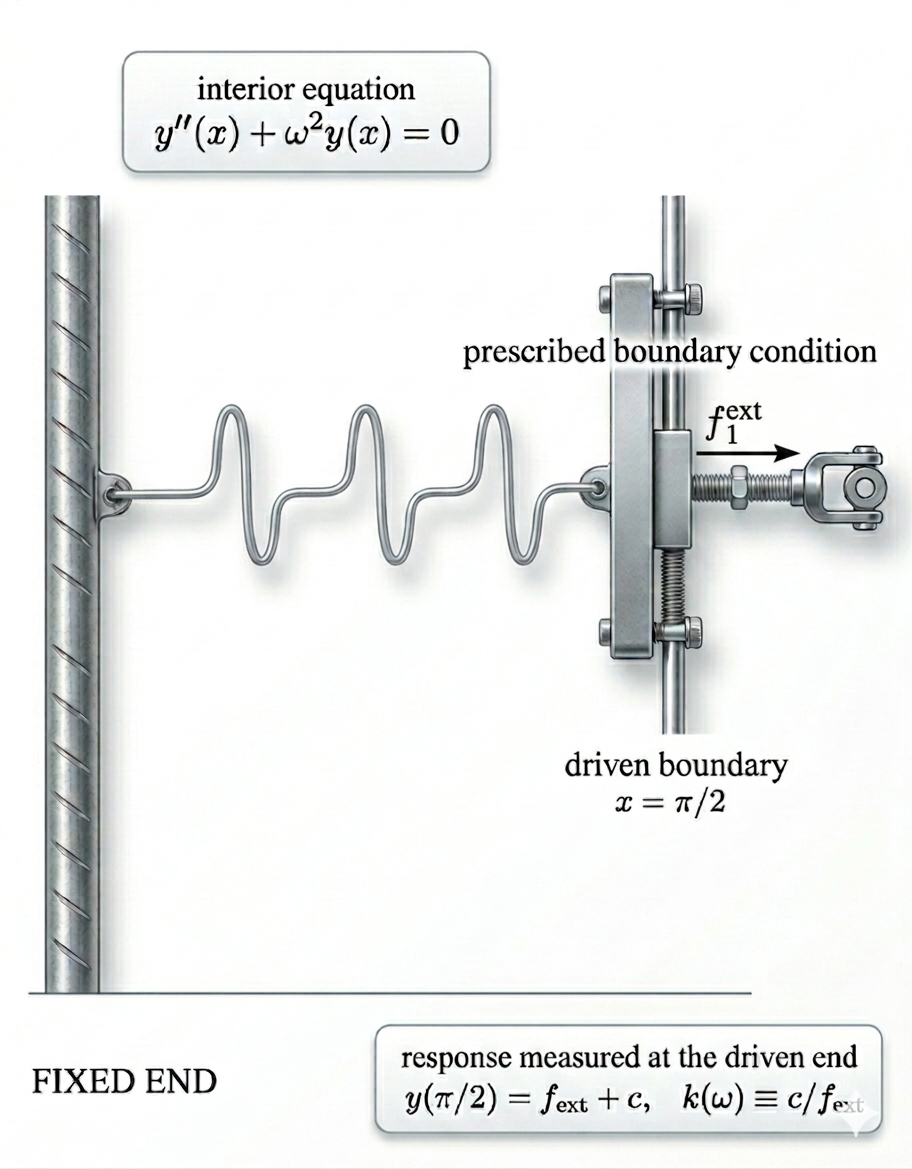}
    \caption{Cartoon depiction of our first toy problem: a string with one end fixed and the other end ``free,'' up to some external forcing of the boundary. }
    \label{fig:toy-problem-cartoon}
\end{figure}

Consider a string of length $\pi/2$, fixed at one end and driven harmonically through a prescribed traction at the other end, as shown through the cartoon in Fig.~\ref{fig:toy-problem-cartoon}. After Fourier transforming in time, the spatial profile $y(x)$ satisfies the boundary-value problem
\begin{align}
    \label{eq:toy-problem-11}
    &y''(x) + \omega^2 y(x) = 0 \,,
\end{align}
with boundary conditions
\begin{align}
    \label{eq:toy-problem-21}
    &y(0) = 0 \,, \qquad y'(\pi/2) = f^{\mathrm{ext}}_1 \,,
\end{align}
where $f^{\mathrm{ext}}_1$ is a constant that characterizes the applied boundary traction (note that $'$ denotes a derivative with respect to $x$ in this section). 
To define a response function, we write the displacement at the driven end as
\begin{align}
    \label{eq:inferred-bc-toy-1}
    y(\pi/2) \equiv f_{\mathrm{ext}} \left[ 1 + k(\omega) \right] \,,
\end{align}
where we have introduced a reference amplitude $f_{\mathrm{ext}}$, and a dimensionless quantity $k(\omega)$ that we can interpret as the string's response to the external driving at the boundary (if we establish a linear connection between the reference amplitude $f_{\mathrm{ext}}$ and the boundary traction $f^{\mathrm{ext}}_1$). In the tidally perturbed neutron-star problem, this response function plays the role of the tidal response function.

The toy problem above is of Sturm-Liouville type in spite of the non-standard boundary conditions. 
Recall that a regular Sturm-Liouville problem is a second-order, linear, eigenvalue problem of the form
\begin{align}
    -\frac{d}{dx}\left[p(x)\frac{dy}{dx}\right] + q(x)y = \lambda s(x)y \,,
\end{align}
posed on a finite interval, with positive-definite functions $p(x)>0$ and $s(x)>0$ in this finite interval, together with linear, homogeneous, self-adjoint boundary conditions that do not depend on the eigenvalue parameter $\lambda$. Since the boundary conditions do not depend on $y$ or on the eigenvalue $\lambda$, and since all other conditions are met, the above problem is of Sturm-Liouville type. 

The \textit{exact} analytical solution to Eq.~\eqref{eq:toy-problem-11} with the boundary conditions of Eq.~\eqref{eq:toy-problem-21} is
\begin{align}\label{eq:y-analytical}
    y(x) = 
    \frac{f^{\mathrm{ext}}_1}{\omega} \frac{\sin(\omega x)}{\cos(\omega \pi/2)}
    \,,
\end{align}
and therefore, we have that the response is 
\begin{align}\label{eq:k-analytical}
    k(\omega) & 
    =\frac{y(\pi/2)}{f_{\mathrm{ext}}} -1\,,
    \\
    &= \frac{f^{\mathrm{ext}}_1}{f_{\mathrm{ext}}} \frac{\tan\left[ \frac{\pi}{2} \omega \right]}{\omega} -1 \,.
\end{align}
The solution of Eq.~\eqref{eq:y-analytical} is exact so it makes no connection to the eigenmodes or the eigenfrequencies of the problem. However, notice that $k(\omega)$ is infinite when $\omega \in 2 \mathbb{Z} + 1$ is an odd integer.
These are precisely the eigenfrequencies of Eq.~\eqref{eq:toy-problem-11} in the absence of any external forces, because then the homogeneous boundary condition would be $y'(\pi/2) = 0$, which would lead to $\cos(\omega \pi/2) = 0$. 

Let us now make the connection to the eigenfunction expansion explicit and relate this problem more cleanly to its Sturm-Liouville structure. To do this, we first choose an \textit{arbitrary function} $F_{\mathrm{ext}}(x)$ that satisfies the inhomogeneous boundary conditions
\begin{align}
\label{eq:boundary-toy-1}
    F_{\mathrm{ext}}(x=0) = 0\,,
    F_{\mathrm{ext}}(\pi/2) = f_{\mathrm{ext}} \,, F_{\mathrm{ext}}'(\pi/2) = f^{\mathrm{ext}}_1 \,.
\end{align}
This function satisfies the above conditions at the boundary, but it \textit{extends} into the interior of the finite domain. 
Once a force $F_{\mathrm{ext}}$ is chosen, we can write Eq.~\eqref{eq:toy-problem-11} in terms of $\tilde{{y}} \equiv {y} - F_{\mathrm{ext}}$
\begin{subequations}\label{eq:toy-problem-forced-sturm-liouville}
\begin{align}
    \label{eq:toy-problem-1-forced-sl}
    &\tilde{{y}}''(x) + \omega^2 \tilde{{y}}(x) = \mathcal{F}_{\mathrm{ext}} \,, \\
    \label{eq:toy-problem-2-forced-sl}
    &\tilde{{y}}(0) = 0 \,, \tilde{{y}}'(\pi/2) = 0 \,,
\end{align}
\end{subequations}
where 
\begin{align}
    \mathcal{F}_{\mathrm{ext}} \equiv - F_{\mathrm{ext}}''(x) - \omega^2 F_{\mathrm{ext}}(x) \,.
\end{align}
After subtracting the arbitrary extension $F_{\rm ext}(x)$, the associated homogeneous problem corresponds to the regular Sturm--Liouville choice $p(x)=1$, $q(x)=0$, $s(x)=1$, and $\lambda=\omega^2$. It is therefore natural to expand $\tilde {y}$ in the corresponding orthogonal eigenfunctions. 

Given this, let us use standard tools from Sturm-Liouville theory to obtain a mode-sum expansion for $\tilde{y}$: 
\begin{align}\label{eq:sl-mode-sum}
    \tilde{{y}}(x) = \sum_{k=0}^{\infty} a_{k}(\omega) \tilde{{y}}_k(x)\,,
\end{align}
where $\tilde{y}_k(x) \equiv \sin[ (2k+1) x]$ are the eigenfunctions of the $\tilde{y}$ problem compatible with the boundary conditions in Eq.~\eqref{eq:toy-problem-2-forced-sl}, and the mode amplitudes $a_k$ are determined by the overlap integral
\begin{align}
    \label{eq:ak-toy-1}
    a_k (\omega) 
    &=
    \frac{4}{\pi \left(\omega^2 - \omega_k^2\right)}
    \int_0^{\pi/2} dx \; \mathcal{F}_{\mathrm{ext}}(x) \; \tilde{{y}}_k (x)
    \,.
\end{align}

Up to this point, the forcing function has been arbitrary. As we show now, this arbitrary function determines the rate of convergence of the mode sum in Eq.~\eqref{eq:sl-mode-sum}.
For comparison, let us use two different external functions:
\begin{subequations}\label{eq:F-choices-toy-model}
\begin{align}
    \label{eq:F_ext_poly_toy_model}
    &F_{\mathrm{ext},\mathrm{poly}}(x) = \left[-\frac{\pi  f^{\mathrm{ext}}_1-4 f_{\mathrm{ext}}}{\pi }\right]x + \frac{2 (\pi  f^{\mathrm{ext}}_1-2 f_{\mathrm{ext}})}{\pi ^2} x^2 \,,\\
    &F_{\mathrm{ext},\mathrm{sine}}(x) = -\frac{f^{\mathrm{ext}}_1}{6} \sin(6x) + \frac{f_{\mathrm{ext}}}{2}[1-\cos(6x)] \,. 
\end{align}
\end{subequations}
The mode amplitudes for each of these functions can be calculated analytically, which we present in Appendix~\ref{appendix:mode-amplitudes-toy-model}.

\begin{figure}[t]
    \centering
    \includegraphics[width=0.99\linewidth]{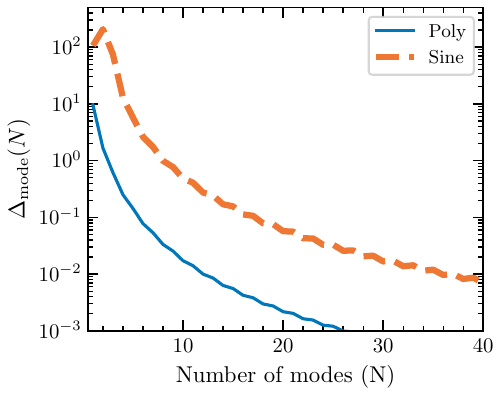}
    \caption{Difference in rate of convergence of mode-sum expression for $k(\omega)$ in the toy model [Eq.~\eqref{eq:kmode-sum}] for different choices of the force [Eq.~\eqref{eq:F-choices-toy-model}]. Observe that the rate of convergence clearly depends on the force function.  
    }
    \label{fig:toy-problem-rate-of-convergence}
\end{figure}

With these expressions, we obtain a mode-sum representation for $k(\omega)$. Using Eq.~\eqref{eq:sl-mode-sum} in Eq.~\eqref{eq:k-analytical}, we find
\begin{align}\label{eq:kmode-sum}
    k(\omega)
    =
    \sum_{k=0}^{\infty} \frac{a_k(\omega)}{f_{\mathrm{ext}}} (-1)^k
    \,.
\end{align}
Numerically, the mode-sum representation for the solution profile inside the star [Eq.~\eqref{eq:sl-mode-sum}] converges to the analytical solution in Eq.~\eqref{eq:y-analytical} for both choices in Eq.~\eqref{eq:F-choices-toy-model}. However, the convergence rate typically differs. To illustrate this difference, let us choose $\omega = 0.9$, $f_{\mathrm{ext}} = 1$ and $f^{\mathrm{ext}}_1=0.1$, to obtain $k(\omega) \approx -0.29847$ from Eq.~\eqref{eq:k-analytical}. In Fig.~\ref{fig:toy-problem-rate-of-convergence}, we assess the rate of convergence for the different choices of $F_{\mathrm{ext}}$ as a function of the number of modes, using the measure
\begin{align}\label{eq:toy-mode-convergence-diagnostic}
    \Delta_{\mathrm{mode}}(N) \equiv \left|1 - \frac{1}{k(\omega)}\sum_{k=0}^{N-1}\frac{a_k(\omega)}{f_{\mathrm{ext}}} (-1)^k\right|\times 100
    \,.
\end{align}
Observe in Fig.~\ref{fig:toy-problem-rate-of-convergence} that the rate of convergence depends strongly on the choice of ``force'' function inside the star. Even though the driving frequency $\omega = 0.9$ is close to the resonant mode $\omega_{0} = 1$ of the system, one needs modes higher than the $N=0$ mode to ensure convergence.

The toy problem discussed above illustrates several important aspects of eigenfunction expansions with non-homogeneous boundary conditions. First, the observables determined by direct integration with the correct boundary conditions are well-behaved and should be regarded as the ultimate ``truth.'' Second, the choice of function to extend the non-homogeneous boundary condition to the interior domain is completely arbitrary, and hence one should not regard the overlap integrals as fundamental objects. That is, in our toy problem, the choice of function $F_{\rm{ext}}(x)$ is arbitrary (up to satisfaction of the boundary conditions of Eq.~\eqref{eq:boundary-toy-1}), and thus, the overlap integrals that give us the mode amplitudes $a_k$ in Eq.~\eqref{eq:ak-toy-1} are not invariant observables. The physical question is not whether a given overlap integral is unique, but whether the tidal response function $k(\omega)$ is robust under reasonable choices of the interior extension. Third, different choices for the arbitrary extension lead to different rates of convergence for the physical observables. Some choices will lead to faster convergence to the true answer, and one expects that, for this problem, all choices will converge if an infinite number of terms are kept. As we will see next, this is only true because of the Sturm-Liouville nature of the toy problem we constructed above.
\subsection{Boundary value problem with frequency dependent boundary conditions}\label{sec:frequency-dependent-sturm-liouville-problem}

Let us now consider a similar toy problem: still a string of length $\pi/2$ that is fixed at one end, but now let the driving through a prescribed traction be frequency dependent. Mathematically, we now have the free oscillator problem of Eq.~\eqref{eq:toy-problem-11}, but the boundary conditions depend on the frequency of the oscillator,
\begin{subequations}\label{eq:toy-problem}
\begin{align}
    \label{eq:toy-problem-2}
    &{y}(0) = 0 \\
    \label{eq:toy-problem-2-bc}
    &{y}'(\pi/2) 
    = \epsilon \, g(\omega) \, {y}(\pi/2) + f^{\mathrm{ext}}_1 - \epsilon \, g(\omega) \, f_{\mathrm{ext}} \,,
\end{align}
\end{subequations}
In the above equations, $f_{\mathrm{ext}}$, $f^{\mathrm{ext}}_1$ and $ \left|\epsilon\right| \ll 1$ are all constants.

The tidal excitation problem for a quasi-circular neutron star binary conceptually maps to this toy problem in the regime $\left|\epsilon\right| \ll 1$. The physical domain $0<x<\pi/2$ in the toy problem corresponds to the radial extent of a neutron star in its rest frame. Equation~\eqref{eq:toy-problem-11} gives the linearized equations of motion of the star, and the nonhomogeneous boundary condition at $x=\pi/2$ captures the boundary condition for tidal excitations. The quantity $\epsilon$ measures the retardation present in the homogeneous solution due to relativistic effects.

The above problem is not of Sturm--Liouville type. The reason is that the spectral parameter, here $\omega$ or equivalently $\lambda=\omega^2$, appears explicitly in the boundary condition. In a typical Sturm--Liouville problem, the boundary conditions are linear, homogeneous, self-adjoint, and independent of the eigenvalue; these properties are what lead to the standard orthogonality and completeness theorems. Here, the frequency-dependent boundary condition spoils that structure in the usual $L^2$ inner product. 
We will soon see how this impacts the solution.

With those preliminaries explained, let us just solve the problem now. One can repeat the analysis carried out in Sec.~\ref{sec:sturm-liouville-problem} to obtain the exact, analytical solution of this second toy problem 
\begin{align}
    \label{eq:y-analytical-2}
    y(x) = 
    \frac{( \epsilon g f_{\mathrm{ext}} - f^{\mathrm{ext}}_1) \sin(\omega x)}{\epsilon g \sin\left(\frac{\pi}{2} \omega\right) - \omega \cos\left(\frac{\pi}{2} \omega\right)}
    \,,
\end{align}
and from this solution, the response function $k(\omega)$ again analytically and exactly: 
\begin{align}
    k(\omega) = \frac{f^{\mathrm{ext}}_1 \sin \left(\frac{\pi  \omega }{2}\right)-f_{\text{ext}} \omega  \cos \left(\frac{\pi  \omega }{2}\right)}{f_{\text{ext}} \omega  \cos \left(\frac{\pi  \omega }{2}\right)
    -
    g(\omega) f_{\text{ext}} \epsilon  \sin \left(\frac{\pi  \omega }{2}\right)}
    \,.
\end{align}
The above expressions do not rely on a small-frequency expansion and are thus valid for any function $g(\omega)$. 

We now analyze Eq.~\eqref{eq:toy-problem-11} with a simplified version of the boundary condition of Eq.~\eqref{eq:toy-problem-2-bc} to show that, in the regime $|\epsilon|\ll 1$, one can introduce eigenfunctions that accurately capture the system's response. In particular, we adopt the boundary conditions of Eq.~\eqref{eq:toy-problem}, but we set $f_{\rm ext} = 0 = f^{\rm ext}_1$, and refer to these as ``homogeneous'' or ``free'' boundary conditions. Consider the differential equation of our toy problem (given in Eq.~\eqref{eq:toy-problem-11}), and multiply it with another solution $\hat{{y}}$ with associated frequency $\hat{\omega}$. Integrate over the domain of the problem to obtain
\begin{align}
       \int_{0}^{\pi/2} dx \, \hat{{y}} \, {y}'' + \omega^2
       \,  \int_{0}^{\pi/2} dx \, \hat{{y}} \, {y} = 0 \,.
\end{align}
Next, integrate the first term by parts to find
\begin{align}
     &\hat{y}(\pi/2) {y}'(\pi/2)
     -
     \hat{{y}}(0) {y}'(0)
     -\int_{0}^{\pi/2} dx \, \hat{{y}}' \, {y}'
     \nonumber\\
     &+
     \omega^2
    \int_{0}^{\pi/2} dx \, \hat{{y}} \, {y} = 0\,.
\end{align}
Now, substitute the free boundary conditions to find 
\begin{align}\label{eq:U-def-toy-problem-2}
    \boldsymbol{U}
    \left[y,\hat{y}\right]
    &\equiv 
    \omega^2
    \int_0^{\pi/2} dx 
    \, y \, \hat{y}
    -
    \int_0^{\pi/2} dx 
    \, y' \, \hat{y}'
    \nonumber\\
    &+
    \epsilon g(\omega) y(\pi/2) \hat{y}(\pi/2)
    =0
    \,,
\end{align}
where $\boldsymbol{U}[\cdot,\cdot]$ is a (not-necessarily symmetric) operator on the space of solutions, which we denote with a boldface font.
If $\epsilon = 0$, this system is of the Sturm-Liouville type, and we can obtain a complete set of basis functions. 
When $\epsilon \neq 0$, it is not clear if eigenfunctions of the system form a basis. 

Let us then assume that the function $g(\omega)$ admits a low-frequency expansion of the form
\begin{align}\label{eq:g-expansion}
    g(\omega) = \omega^2 \left[1 + \mathcal{O}(\omega^4) \right]
    \,,
\end{align}
and carry out a perturbative analysis. 
If $|\epsilon| \ll 1$, we can show that the \textit{eigenfunctions are orthogonal with respect to a symmetric operator}. To see this, substitute the low-frequency approximation of $g(\omega)$ in the above equation and consider the difference in operators
\begin{align}\label{eq:orthogonality-condition}
    &\boldsymbol{U}
    \left[y,\hat{y}\right] - 
    \boldsymbol{U}
    \left[\hat{y},{y}\right]
    \equiv
    \left(\omega^2 - \hat{\omega}^2 \right)
    \boldsymbol{O}\left[y,\hat{y}\right]
    =
    0
    \,,
\end{align}
where
\begin{align}\label{eq:O-operator}
    \boldsymbol{O}\left[y,\hat{y}\right]
    \equiv 
    \int_0^{\pi/2} \hat{y} \, y \, dx + \epsilon \, y(\pi/2) \, \hat{y}(\pi/2) 
    \,
\end{align}
is a new \textit{symmetric} operator on the space of solutions. Equation~\eqref{eq:orthogonality-condition} tells us that eigenfunctions with different frequencies are orthogonal with respect to the operator $\boldsymbol{O}$.
\textit{This does not mean that the eigenfunctions form a complete set.}
In the PN literature~\cite{gittins2025perturbationtheorypostnewtonianneutron,HegadeKR:2025qwj}, we typically assume this is the case. But formally speaking, this need not be the case because we need to ensure that the operators are well-defined on a function space with a positive definite inner product~\cite{walter1973regular,Fulton1977}.

For the moment, however, let us ignore this technical point and assume the eigenfunctions form a complete set to see what happens. Below, we make this assumption to develop a \textit{formal} mode-sum approximation to $k(\omega)$ by repeating the steps leading to Eq.~\eqref{eq:kmode-sum} in the previous subsection. First, we extend the ``force'' inside the star by introducing an arbitrary function $F(x)$ that extends inside the domain of the problem and satisfies
\begin{align}
    F_{\mathrm{ext}}(\pi/2) = f_{\mathrm{ext}} \,,
    F_{\mathrm{ext}}'(\pi/2) = f^{\mathrm{ext}}_1 \,,
    F_{\mathrm{ext}}(0) = 0 \,.
\end{align}
As in the previous toy problem, let us define $\tilde{y} \equiv y - F_{\rm ext}$, and define
\begin{align}
    &\mathcal{F}_{\mathrm{ext}} \equiv - \omega^2 F_{\rm ext}(x) - F_{\rm ext}''(x) \,
\end{align}
to get
\begin{align}
    \tilde{y}'' + \omega^2 \tilde{y}  = \mathcal{F}_{\mathrm{ext}} \,.
\end{align}
Multiply the above equation with a solution $\hat{\tilde{y}}_p$ to the homogeneous problem and integrate over the domain to find
\begin{align}
    \omega^2 \int_{0}^{\pi/2} dx \,
    \tilde{y} \, \hat{\tilde{y}}_p + 
    \int_{0}^{\pi/2} dx 
    \, \tilde{y}'' \, \hat{\tilde{y}}_p
    = 
    \int_{0}^{\pi/2} dx \, \mathcal{F}_{\mathrm{ext}} \hat{\tilde{y}}_p
    \,.
\end{align}
Integrate the second term by parts and use the homogeneous boundary conditions to obtain
\begin{align}\label{eq:tidal-overlap-1-toy-model}
    \omega^2
    &\int_0^{\pi/2} dx 
    \tilde{y} \hat{\tilde{y}}_p
    -
    \int_0^{\pi/2} dx 
    \tilde{y}' \hat{\tilde{y}}'_p
    +
    \epsilon g(\omega) \tilde{y}(\pi/2) \hat{\tilde{y}}_p(\pi/2)
    \nonumber\\
    &=
    \int_0^{\pi/2} \mathcal{F}_{\mathrm{ext}} \hat{\tilde{y}}_p dx \,.
\end{align}

Now, we \textit{assume} that we can use the eigenfunctions as a basis to expand $\tilde{y}$
\begin{align}
    \tilde{y} = \sum_{s} a_{s} \tilde{y}_{s} \,.
\end{align}
Substitute this in Eq.~\eqref{eq:tidal-overlap-1-toy-model} to find
\begin{align}\label{eq:overlap-toy-problem-2}
    &
    \sum_{s}
    a_s
    \bigg[
    \omega^2
    \int_0^{\pi/2} dx 
    \, \tilde{y}_s \, \hat{\tilde{y}}_p
    -
    \int_0^{\pi/2} dx \, 
    \tilde{y}'_s \, \hat{\tilde{y}}'_p
    \nonumber\\
    &+
    \epsilon g(\omega) \, \tilde{y}_s(\pi/2) \hat{\tilde{y}}_p(\pi/2)
    \bigg]
    =
    \int_0^{\pi/2} \mathcal{F}_{\mathrm{ext}} \, \hat{\tilde{y}}_p \, dx
    \,.
\end{align}
We note from Eq.~\eqref{eq:U-def-toy-problem-2} that
\begin{align}
    &
    a_s
    \boldsymbol{U}\left[\tilde{y}_s, \hat{\tilde{y}}_p\right]
    =
    a_s
    \bigg[
    \omega^2_s
    \int_0^{\pi/2} dx 
    \, \tilde{y}_s \, \hat{\tilde{y}}_p
    -
    \int_0^{\pi/2} dx 
    \, \tilde{y}_s' \, \hat{\tilde{y}}_p'
    \nonumber\\
    &+
    \epsilon g(\omega_s) \tilde{y}_s(\pi/2) \hat{\tilde{y}}_p(\pi/2)
    \bigg]
    =0
\end{align}
Subtract this equation from the term inside the square brackets in Eq.~\eqref{eq:overlap-toy-problem-2} and simplify to obtain
\begin{align}
    &
    \sum_{s}
    a_s
    \bigg\{
    (\omega^2 - \omega_s^2)
    \int_0^{\pi/2} dx \,
    \tilde{y}_s \,\hat{\tilde{y}}_p
    \nonumber\\
    &+
    \epsilon \left[g(\omega) - g(\omega_s) \right]
    \tilde{y}_s(\pi/2) \hat{\tilde{y}}_p(\pi/2)
    \bigg\}
    =
    \int_0^{\pi/2} \mathcal{F}_{\mathrm{ext}} \,\hat{\tilde{y}}_p \, dx
    \,.
\end{align}

We can simplify this equation further by substituting the low-frequency boundary conditions for $g(\omega) - g(\omega_s)$ to find
\begin{align}
    \sum_{s}
    a_s(\omega^2 - \omega_s^2)
    \boldsymbol{O}\left[\tilde{y}_s, \hat{\tilde{y}}_p\right]
    =
    \int_0^{\pi/2} \mathcal{F}_{\mathrm{ext}} \,\hat{\tilde{y}}_p \,dx
    \,.
\end{align}
Using the orthogonality condition from Eq.~\eqref{eq:orthogonality-condition}, we can obtain the mode amplitude 
\begin{align}
    &a_s = \frac{I_s}{\left(\omega^2 - \omega_s^2\right) \boldsymbol{O}[\tilde{y}_s,\tilde{y}_s]}
    \,, \\
\end{align}
in terms of the overlap integral 
\begin{align}
    &I_s \equiv \int_0^{\pi/2} \mathcal{F}_{\mathrm{ext}} \tilde{y}_s dx\,.
\end{align}
The amplitudes from the above equation can be used to obtain the mode-sum approximation to $k(\omega)$
\begin{align}
    k(\omega) 
    &=
    \frac{y(\pi/2) - f_{\mathrm{ext}}}{f_{\mathrm{ext}}}
    =
    \frac{\tilde{y}(\pi/2)}{f_{\mathrm{ext}}}
    \nonumber\\
    &=
    \sum_{k=0}^{\infty} \frac{a_k(\omega)}{f_{\mathrm{ext}}} \sin\left[ \frac{\pi}{2} \omega_k \right]
    \,,
\end{align}
where we have used the fact that the eigenfunctions satisfy $\tilde{y}_k(\pi/2) = \sin(\omega_k \pi/2)$. 
The eigenfrequencies $\omega_k$ are determined by using the low-frequency expansion of the boundary conditions. These eigenfrequencies map to the system's true physical eigenfrequencies only for the first few values, since we use a small-frequency expansion of the boundary conditions. 

\begin{figure*}[tbh]
    \centering
    \includegraphics[width=0.99\columnwidth]{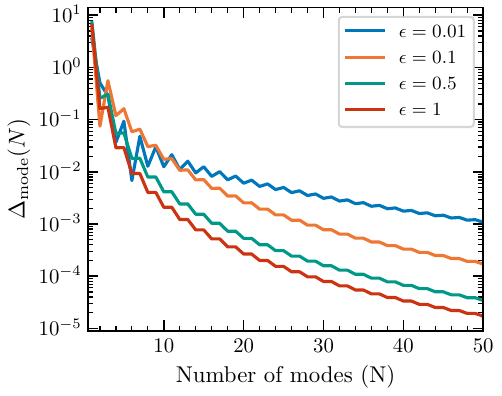}
    \includegraphics[width=0.99\columnwidth]{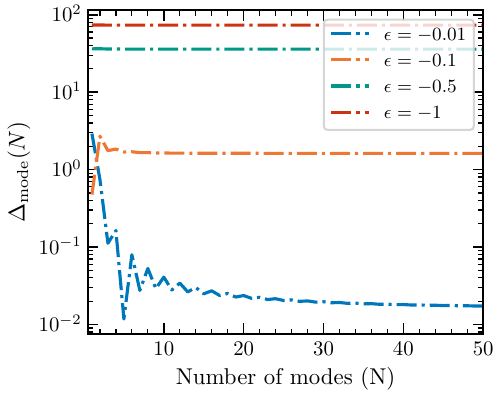}
    \caption{Difference in rate of convergence of the mode sum for the toy model discussed in Sec.~\ref{sec:frequency-dependent-sturm-liouville-problem}. Observe that for all values of $\epsilon >0$ (left panel) the mode-sum converges. However, for $\epsilon<0$ (right panel), the mode sum never converges no matter how small $|\epsilon|$ is.}
    \label{fig:toy-model-plus-minus}
\end{figure*}
In Fig.~\ref{fig:toy-model-plus-minus}, we assess the rate of convergence of the mode sum as a function of the number of modes, using the diagnostic from Eq.~\eqref{eq:toy-mode-convergence-diagnostic}. For this test, we choose $F = F_{\mathrm{ext,poly}}$ for simplicity [see Eq.~\eqref{eq:F_ext_poly_toy_model}], and we explore different values of $\epsilon$, fixing $\omega = 0.01$, $f_{\mathrm{ext}} = 1$, and $f^{\mathrm{ext}}_1 = 0.1$. 
Observe that the mode-sum converges for all positive values of $\epsilon$ (left panel) while it does not converge exactly for any negative value of $\epsilon$ (right panel). This is not a numerical artifact of the toy problem: it is a real mathematical result.

Why do we lose convergence? If we go back and look at our assumptions, we see that we proved that the modes are orthogonal with respect to $\boldsymbol{O}$ [Eq.~\eqref{eq:O-operator}] and then assumed that the modes form a complete set. 
Unfortunately, \textit{orthogonality is not enough to prove completeness}. Intuitively, we can understand this by noticing that when $\hat{y}=y$, $\boldsymbol{O}$ [Eq.~\eqref{eq:O-operator}] is only positive definite when $\epsilon>0$. When $\epsilon<0$, the operator can be negative for certain functions, and therefore, the modes do not necessarily form a complete set, see~\cite{walter1973regular,Fulton1977} for rigorous mathematical discussion. This reminds us of an important lesson: the completeness of a basis requires not just that the eigenfunctions be orthogonal with respect to some operator, but that the operator itself be positive definite.  

As we demonstrate below, the tidal excitation problem for PN and relativistic stars falls in the ``$\epsilon<0$ regime'' of our toy problem, with ``increasing $|\epsilon|$'' for larger compactness (more relativistic) stars. Fortunately, even for highly relativistic stars, we expect the problem to map to the toy problem in the $|\epsilon| \sim 0.1$ regime, and hence we do not expect a completely non-convergent mode-sum approximation, as seen in the red and orange curves in the right panel of Fig.~\ref{fig:toy-model-plus-minus}.
\section{Relativistic Tidal response theory}\label{sec:response-theory}

In this section, we describe the equations governing the non-radial oscillations of a non-rotating neutron star and present the main technical details on the formulation of the tidal overlap integral in general relativity.
The organization of this section is as follows:
In Sec.~\ref{sec:background}, we present the Tolman-Oppenheimer-Volkoff (TOV) equations.
Section~\ref{sec:non-radial-pert} presents the master equation for non-radial perturbations of a neutron star.
We then discuss the solution to the master equation outside the star in Sec.~\ref{sec:exterior-boundary-cond}.
We derive the operator form of the linear perturbation problem in Sec.~\ref{sec:action-principle} and discuss the constraints on the tidal field in Sec.~\ref{sec:constraints-on-tide}.
Section~\ref{sec:mode-amplitude-evolution} discusses the evolution of the mode amplitude due to the tidal field.
In Sec.~\ref{sec:tidal-overlap}, we present the relativistic generalization of the tidal overlap integral.
Section~\ref{sec:summary-and-num} contains a summary of the results and presents a routine for numerical integration of the equations. 

The notation we follow in this section is as follows: 
We consider linearized perturbations to a spherically-symmetric (TOV) star in Regge-Wheeler gauge with Schwarzschild-like coordinates $\left(t,r,\theta,\phi\right)$.
Capital Latin indices $A,B\ldots$ are used to denote the coordinates $\left(\theta,\phi\right)$.
The time dependence of all the perturbed quantities is Fourier transformed using the replacement $\partial_t \to -i \omega$.
Scalar, vector, and tensor polar spherical harmonics are denoted by $Y_{\ell m}$, $E_A^{\ell m }$, and $Z_{AB}^{\ell m }$, respectively.
We only consider polar perturbations in this article.
We set $u^{\mu}$ to be the fluid four velocity, $p$ to be the pressure, $n$ to be the baryon number density,
$\rho = m n$ to be the rest mass energy density, also known as the baryon (mass) density, and $e$ to be the total energy density.
Finally, we use $\delta$ and $\Delta$ to denote the Eulerian and Lagrangian fluid perturbations, respectively.
\subsection{Background metric}\label{sec:background}
The Einstein field equations with a perfect fluid matter source are given by
\begin{subequations}
\begin{align}
    &G_{\mu \nu} = 8 \pi T_{\mu \nu} \,,\\
    & \nabla_{\mu} T^{\mu \nu} = 0\,,
\end{align}
\end{subequations}
where
\begin{align}
    &T_{\mu \nu} = \left(e + p \right) u_{\mu} u_{\nu} + p g_{\mu \nu} \,.
\end{align}
The background spacetime is static and spherically symmetric with line element
\begin{align}\label{eq:background-metric}
    ds^2 = -e^{\nu(r)} dt^2
    +
    e^{\lambda(r)} dr^2
    +
    r^2 
    d \Omega^2
\end{align}
where $d \Omega^2 = d \theta^2 + \sin^2(\theta) d \phi^2$ is the line-element on the 2-sphere.
The background metric and fluid variables satisfy the TOV equations
\begin{subequations}\label{eq:TOV-equations}
\begin{align}
    \label{eq:lambda-equation-TOV}
    \lambda'(r) &= \frac{1- e^{\lambda(r)} + 8 \pi r^2 e^{\lambda(r)} e}{r} \,,\\
    \label{eq:nu-equation-TOV}
    \nu'(r) &= \frac{-1+e^{\lambda }+8 e^{\lambda } \pi  r^2 p}{r} \,,\\
    \label{eq:p-equation-TOV}
    p'(r) &= -\frac{(e+p) \left(-1+e^{\lambda }+8 e^{\lambda } \pi  r^2 p\right)}{2 r }
    \,.
\end{align}
\end{subequations}
Given an EoS $p(e)$ and the central pressure $p(r=0)$, one can integrate the above equations to obtain the background spacetime and fluid variables.
From hereon, we denote the total mass of the star by $M$, the stellar radius by $R$, and the stellar compactness by $C = M/R$.
\subsection{Non-radial perturbations}\label{sec:non-radial-pert}
In this section, we present the master equations governing the non-radial oscillations of a neutron star.
We consider non-radial polar perturbations of the background TOV metric in the Regge-Wheeler gauge. We decompose the linearly perturbed metric components into spherical harmonics $Y_{\ell m}$, so that the line element reads  
\begin{align}\label{eq:metric-polar-pert}
    ds^2 &= -e^{\nu(r)} \left(1 - 2 H(r) e^{-i\omega t}  r^{\ell} Y_{\ell m}\right) dt^2
    \nonumber\\
    &- 
    2 i H_1(r) e^{-i\omega t}  r^{\ell} Y_{\ell m} dt dr
    \nonumber\\
    &+
    e^{\lambda(r)}
    \left(1 + 2 H_2(r) e^{-i\omega t}  r^{\ell} Y_{\ell m }\right) dr^2
    \nonumber\\
    &+
    r^2 \left(
    1 - K(r) e^{-i\omega t} r^{\ell} Y_{\ell m }
    \right)
    d \Omega^2
    \,,
\end{align}
where the purely radial functions $(H(r), H_1(r), H_2(r),K(r))$ are to be determined from the linearized field equations. 
To simplify the fluid perturbations, we introduce the Lagrangian displacement vector $\xi^{\mu}$. The non-zero components of the Lagrangian displacement vector are parameterized as
\begin{subequations}
\begin{align}\label{eq:lagrangian-displacement-vectors}
    \xi^r &= W(r) e^{-i\omega t} r^{\ell - 1} e^{-\lambda/2} Y_{\ell m} \,,\\
    \xi^A &= - V(r) e^{-i\omega t} r^{\ell-2 } E^{A}_{\ell m} \,,
\end{align}
\end{subequations}
where $(W(r),V(r))$ are purely radial functions to be determined from the linearized field equations.

Standard techniques in Lagrangian perturbation theory (see Chapter 7 of~\cite{FN-book}) allow one to obtain the Lagrangian perturbations to the rest mass density, pressure, energy density, and the four-velocity in terms of the Lagrangian displacement vector and the metric perturbations:

\begin{subequations}
\label{eq:fluid-perts}
\begin{align}
    &\frac{\Delta \rho}{\rho} = -\frac{1}{2} \gamma^{\mu \nu} \Delta g_{\mu\nu} \,, \\
    &\frac{\Delta p}{p} = \Gamma \frac{\Delta \rho}{\rho}\,, \\
    &\frac{\Delta e}{e + p} = \frac{\Delta \rho}{\rho} \,, \\
    &\Delta u^{\alpha} = \frac{1}{2} u^{\alpha} u^{\beta} u^{\gamma} \Delta g_{\beta \gamma} \,,
\end{align}
\end{subequations}
where $\gamma_{\mu \nu} = u_{\mu} u_{\nu} + g_{\mu \nu}$ is a projection tensor, and $\Gamma$ is the adiabatic index.

The linearized Einstein equations can be reduced to a set of ordinary differential equations for the variables $\boldsymbol{Y} = (\tilde{H}_{1}, K, W, X$)~\cite{1983ApJS...53...73L,1985ApJ...292...12D} 
where
\begin{subequations}
\begin{align}
    \tilde{H}_1 &\equiv \frac{H_{1}}{r \omega} \,,\\
    X &\equiv \frac{\Delta p }{r^{\ell} (e + p)} e^{\nu/2} \,,
\end{align}
\end{subequations}
where $X$ can be written in terms of the metric perturbation through Eq.~\eqref{eq:fluid-perts}. The linearized Einstein equations give algebraic relations for the other variables $(H,H_2,V)$ in terms of $\boldsymbol{Y}$ (see Appendix~\ref{appendix:DI-eom}). Schematically, the master equations are a system of first order differential equations of the form
\begin{align}\label{eq:master-equation-matrix}
    \boldsymbol{Y}' &= \boldsymbol{A} \cdot \boldsymbol{Y}\,,
\end{align}
where Appendix~\ref{appendix:DI-eom} presents explicit expressions. The master equations can be used to reconstruct all metric and fluid variables, but they are not unique; one can always use a different parameterization or a different gauge to obtain different but physically-equivalent master equations (see e.g.~\cite{1983ApJS...53...73L,Lindblom_1997,FN-book,HegadeKR:2024agt} for other approaches).

Outside the star, the background spacetime is the Schwarzschild metric, and the master equations reduce to the Zerilli-Moncrief equation~\cite{Martel_2005,Berti_2009}
\begin{align}
    \frac{d^2}{d r_{\star}^2} Z + \left( \omega^2 - V_{\mathrm{ZM}} \right) Z = 0\,,
\end{align}
where $r_{\star} = r + 2 M \log(r/2M - 1)$ is the tortoise-coordinate,
\begin{align}
    Z\equiv \frac{2 r^\ell (-2 M+r) H_1 \omega^{-1}}{6 M+\left(-2+\ell+\ell^2\right) r}+\frac{2 r^{2+\ell} K}{6 M+\left(-2+\ell+\ell^2\right) r}
\end{align}
is the Zerilli-Moncrief master function, and
\begin{widetext}
\begin{align}
    V_{\mathrm{ZM}} \equiv 
    \frac{(r-2M) \left(36 \left(\ell^2+\ell-2\right) M^2 r+6 \left(\ell^2+\ell-2\right)^2 M r^2+\ell (\ell+1) \left(\ell^2+\ell-2\right)^2 r^3+72 M^3\right)}{r^4 \left(\left(\ell^2+\ell-2\right) r+6 M\right)^2}
    \,
\end{align}
\end{widetext}
is the Zerilli-Moncrief potential. All the metric functions outside the star can be reconstructed using the Zerilli-Moncrief function (see Appendix~\ref{appendix:DI-eom}).

\subsection{Solution to the master equation outside the star}\label{sec:exterior-boundary-cond}
In this section, we recall the solution to the master equation [Eq.~\eqref{eq:master-equation-matrix}] outside the star obtained in detail in Sec. IVB of~\cite{HegadeKR:2024agt}.
The solution is obtained by solving the Zerilli-Moncrief equation outside the star in a small-frequency ($ \varepsilon \equiv M \omega \ll 1$) expansion with the normalization discussed in~\cite{HegadeKR:2024agt}.
For the analysis below, we only need the schematic form of the solution to $Z$ outside the star:
\begin{subequations}\label{eq:Z-out-schematic}
\begin{align}
    Z &= Z_{T} + Z_{I} \,,\\
    Z_{T}&=\frac{4 \pi d_{\ell m}(\omega) }{2 \ell +1}
    \left[ z_{0,T}(r) + z_{1,T}(r) \omega^2  \right]
    + 
    \mathcal{O}\left(\varepsilon^4\right)\,,\\
    Z_{I}&=
    \frac{4 \pi I_{\ell m}(\omega)}{ (2 \ell +1)}
    \left[ z_{0,I}(r) + z_{1,I}(r) \omega^2  \right]
    + 
    \mathcal{O}\left(\varepsilon^4\right)
    \,,
\end{align}
\end{subequations}
where, $Z_{T}$ is the contribution from the tidal field, $Z_{I}$ is the contribution from the multipole moment of the star, $d_{\ell m}(\omega)$ is the tidal moment of the external source, $I_{\ell m}(\omega)$ is the multipole moment of the neutron star.
The supplementary \texttt{Mathematica} notebook lists the coefficients $z_{i,T/I}$ appearing in the above equation, along with expressions for other metric functions outside the star and for $\ell=2$ moment.
\subsection{Operator formulation for the perturbations}\label{sec:action-principle}
In this section, we derive an operator formulation for non-radial perturbations.
Let $\boldsymbol{\boldsymbol{y}}$ and $\hat{\boldsymbol{\boldsymbol{y}}}$ be two abstract vectors,
\begin{subequations}
\begin{align}
    \boldsymbol{\boldsymbol{y}} &\equiv e^{-i \omega t} Y_{\ell m}\left(H, H_1, H_2,K, V, W\right) \,,\\
    \hat{\boldsymbol{\boldsymbol{y}}} &\equiv  e^{-i \hat{\omega} t} Y_{\ell m} \left(\hat{H}, \hat{H}_1, \hat{H}_2,\hat{K}, \hat{V}, \hat{W}\right) \,,
\end{align}
\end{subequations}
Friedman and Schutz~\cite{FS_stability_rel,Friedmann-rotating-stars,FN-book} 
showed that the following identity is valid for any two arbitrary abstract vectors
\begin{align}\label{eq:FS-identity}
    \hat{\xi}_{\beta} \boldsymbol{E}^{\beta}[\boldsymbol{y}] + \frac{\hat{h}_{\alpha \beta} \boldsymbol{E}^{\alpha \beta}[\boldsymbol{y}] }{16 \pi} = - \pmb{\mathscr{L}}[\hat{\boldsymbol{y}}, \boldsymbol{y}] + \nabla_{\beta} \boldsymbol{\Theta}^{\beta}[\hat{\boldsymbol{y}}, \boldsymbol{y}]
    \,,
\end{align}
where $\hat{\xi}^{\beta}$ is the Lagrangian displacement vector, $\hat{h}_{\alpha \beta}$ is the metric perturbation of the ``hatted'' variables, $\pmb{\mathscr{L}}[\hat{\boldsymbol{y}},\boldsymbol{y}] = \pmb{\mathscr{L}}[\boldsymbol{y},\hat{\boldsymbol{y}}]$ is a symmetric operator, $\boldsymbol{\Theta}$ is a boundary term, $\boldsymbol{E}^{\alpha \beta}[\boldsymbol{y}] $ denotes the linearized Einstein equations and $\boldsymbol{E}^{\alpha}[\boldsymbol{y}]$ are the linearized stress energy conservation equations.
The explicit expressions for the operators $\pmb{\mathscr{L}}$ and $\boldsymbol{\Theta}$ can be found in Chapter 7 of~\cite{FN-book} and in the appendix of Paper I. 
Notice that the validity of the above equation \textit{does not} rely on the use of the linearized field equations for vectors $\boldsymbol{\boldsymbol{y}}$ or $\hat{\boldsymbol{\boldsymbol{y}}}$, but it does require that the background satisfy the Einstein equations at background order.
Using Eq.~\eqref{eq:FS-identity}, one can construct an action principle, see Chapter 7 of~\cite{FN-book}.

We use Eq.~\eqref{eq:FS-identity} as a starting point to derive our operator formulation.
To do this, we demand that the abstract vectors we consider satisfy the following conditions:
\begin{itemize}
    \item[i)] $\boldsymbol{\boldsymbol{y}}$ satisfies the linearized equations inside and outside the star.
    \item [ii)] $\hat{\boldsymbol{\boldsymbol{y}}}$ satisfies the linearized Einstein equations \textit{outside} the star and the linearized Hamiltonian and momentum constraints inside the star.
    \item  [iii)] The abstract vectors satisfy the boundary condition $Z = Z_{I}$ [Eq.~\eqref{eq:Z-out-schematic}] at the surface of the star.
\end{itemize}
The first and second conditions allow an arbitrary vector $\hat{\boldsymbol{\boldsymbol{y}}}$, but restrict the solution set of the operator formulation to the solutions satisfying the initial value constraints inside the star.
The third condition provides the right boundary conditions for the perturbations outside the star.

We now perform a $3+1$ split of the metric perturbations and the linearized Einstein equations.
We first split the metric perturbation $h_{\alpha \beta}$ as
\begin{align}
    h_{\alpha \beta} &\equiv u_{\alpha} u_{\beta} \mathfrak{h} + i \omega u_{\alpha} \mathfrak{h}_{\beta} + i \omega u_{\beta} \mathfrak{h}_{\alpha} + \mathfrak{h}_{\alpha \beta}
    \,,
\end{align}
with a similar decomposition for the $\hat{h}_{\mu \nu}$.
In Regge-Wheeler gauge, we can use Eq.~\eqref{eq:metric-polar-pert}, to obtain
\begin{subequations}
\begin{align}
    &\mathfrak{h} = 2 H r^{\ell} \,,\\
    &i \omega \mathfrak{h}_{\alpha} = (0,-i H_{1} r^{\ell} e^{-\nu/2}, 0, 0) \,,\\
    &\mathfrak{h}_{\alpha \beta} = 
    \mathrm{diag}
    \left[0,2 H_2 e^{\lambda} , - K r^{2}, - K r^{2} \sin^2(\theta)  \right] r^{\ell}
    \,.
\end{align}
\end{subequations}
Notice that we have suppressed the $e^{- i \omega t} Y_{\ell m}$ factor in the above equation. 
We shall do the same in the following to reduce the length of our expressions.

Define the linearized Hamiltonian and momentum constraint operators of the system
\begin{align}
    \boldsymbol{H}[\boldsymbol{y}] \equiv \boldsymbol{E}_{\mu \nu}[\boldsymbol{y}] u^{\mu} u^{\nu} \,, \quad 
    (- i \omega) \boldsymbol{P}_{\rho}[\boldsymbol{y}] \equiv \boldsymbol{E}_{\mu \nu}[\boldsymbol{y}] u^{\mu} \gamma^{\nu}_{\rho} \,,
\end{align}
where $\gamma^{\nu}_{\rho} = \delta^{\nu}_{\rho} + u^{\nu} u_{\rho}$ is the projector onto spatial hypersurfaces.
By integrating the Hamiltonian constraint, the momentum constraint, and the right-hand side of Eq.~\eqref{eq:FS-identity} by parts, one can show that the following identities are true
\begin{subequations}
\begin{align}
    \label{eq:operator-form-hamiltonian-general-gauge}
    &\int d \Omega 
    \frac{e^{\nu/2}}{16\pi}\hat{\mathfrak{h}} \boldsymbol{H} = 
    \pmb{\mathcal{A}}_{H}[\hat{\boldsymbol{y}},\boldsymbol{y}] + \frac{1}{e^{\lambda(r)/2} r^2}\frac{d}{dr} \pmb{\mathcal{R}}_{H}[\hat{\boldsymbol{y}},\boldsymbol{y}] \,,\\
    \label{eq:operator-form-momentum-general-gauge}
    &\int d \Omega
    \frac{e^{\nu/2}}{8\pi}\hat{\mathfrak{h}}^{\rho} \boldsymbol{P}_{\rho} = 
    \pmb{\mathcal{A}}_{P}[\hat{\boldsymbol{y}},\boldsymbol{y}] + \frac{1}{e^{\lambda(r)/2} r^2}\frac{d}{dr} \pmb{\mathcal{R}}_{P}[\hat{\boldsymbol{y}},\boldsymbol{y}] \,, \\
    \label{eq:Lagrangian-integrated-by-parts-1}
    &\int d \Omega
    e^{\nu/2}
    \bigg[
    \hat{\xi}_{\beta} \boldsymbol{E}^{\beta}[\boldsymbol{y}] + \frac{\hat{h}_{\alpha \beta} \boldsymbol{E}^{\alpha \beta}[\boldsymbol{y}] }{16 \pi} 
    \nonumber\\
    &-
    \frac{\hat{\omega} \omega \hat{\mathfrak{h}}_{\rho}}{8\pi} \boldsymbol{P}^{\rho}
    +
    \frac{\omega^2}{8\pi}
    \hat{\boldsymbol{P}}^{\rho}
    \mathfrak{h}_{\rho}
    \bigg]
    =
    -\omega^2
    \boldsymbol{O}_{0,\DI}[\hat{\boldsymbol{y}},\boldsymbol{y}] +
    \boldsymbol{O}_{1}[\hat{\boldsymbol{y}},\boldsymbol{y}] \nonumber\\
    &\hspace{2cm}
    +
    \frac{1}{e^{\lambda(r)/2} r^2}\frac{d}{dr} \pmb{\mathcal{R}}[\hat{\boldsymbol{y}},\boldsymbol{y}]
    \,,
\end{align}
\end{subequations}
where the subscript \DI\, stands for Detweiler-Ipser who derived an identity equivalent to Eq.~\eqref{eq:Lagrangian-integrated-by-parts-1} in~\cite{1973ApJ...185..685D}.
However, our operators only reduce to the Detweiler-Ipser operators on shell, since they used certain components of the Einstein equations (such as the $(r,\theta)$ component) to simplify some of their expressions. 
The explicit expressions for the operators are listed in Appendix~\ref{appendix:operators-proof}. 

Define the operator 
\begin{widetext}
\begin{align}\label{eq:integration-v1}
    \boldsymbol{U}[\hat{\boldsymbol{y}}, \boldsymbol{y}] 
    &\equiv
    \int r^2 e^{\lambda/2} dr d \Omega
    e^{\nu(r)/2}
    \bigg[
    \hat{\xi}_{\beta} \boldsymbol{E}^{\beta}[\boldsymbol{y}] + \frac{\hat{h}_{\alpha \beta} \boldsymbol{E}^{\alpha \beta}[\boldsymbol{y}] }{16 \pi} 
    -
    \frac{\hat{\omega} \omega}{8\pi} \boldsymbol{P}^{\rho}  \hat{\mathfrak{h}}_{\rho}
    +
    \frac{\omega^2}{8\pi}
    \hat{\boldsymbol{P}}^{\rho}
    \mathfrak{h}_{\rho}
    \bigg]
    \,, \nonumber\\
    &=
    -\omega^2 \int_{r=0}^{R} r^2 e^{\lambda/2} dr\boldsymbol{O}_{0,\DI}[\hat{\boldsymbol{y}},\boldsymbol{y}] 
    +
    \int_{r=0}^{R} r^2 e^{\lambda/2} dr
    \boldsymbol{O}_{1}[\hat{\boldsymbol{y}},\boldsymbol{y}] 
    +
    \bigg.
    \pmb{\mathscr{R}}\left[\hat{\boldsymbol{y}},\boldsymbol{y}\right] 
    \bigg|_{r=R}\,,
\end{align}
and repeat the analysis from Sec.~\ref{sec:frequency-dependent-sturm-liouville-problem} to show that the linearized solutions are eigenvectors of a symmetric operator.
Consider two solutions to the linearized equations with frequencies $\omega$ and $\hat{\omega}$. For these solutions, construct the operator 
\begin{align}\label{eq:E-equation-2}
     \boldsymbol{E}[\hat{\boldsymbol{y}}, \boldsymbol{y}]
     \equiv\boldsymbol{U}[\hat{\boldsymbol{y}}, \boldsymbol{y}]
    -
    \boldsymbol{U}[ \boldsymbol{y}, \hat{\boldsymbol{y}}]
    =
    - (\omega^2 -\hat{\omega}^2) 
    \bigg[\boldsymbol{J}_0[\hat{\boldsymbol{y}},\boldsymbol{y}] 
    -
    \frac{1}{(\omega^2 -\hat{\omega}^2) }
    \bigg.
    \bigg(
    \pmb{\mathscr{R}}\left[\hat{\boldsymbol{y}},\boldsymbol{y}\right] 
    -
    \pmb{\mathscr{R}}\left[\boldsymbol{y}, \hat{\boldsymbol{y}}\right] 
    \bigg)
    \bigg|_{r=R}
    \bigg]
    \,,
\end{align}
where
\begin{align}\label{eq:boldsymbol-I0-I1}
    &\boldsymbol{J}_0[\hat{\boldsymbol{y}},\boldsymbol{y}] \equiv \int_{r=0}^{R} r^2 e^{\lambda/2} \, \boldsymbol{O}_{0,\DI}[\hat{\boldsymbol{y}},\boldsymbol{y}]  dr
    \,.
\end{align}
\end{widetext}
Inspecting Eqs.~\eqref{eq:E-equation-2} and \eqref{eq:boldsymbol-I0-I1}, we see that $\boldsymbol{J}_0$ is a symmetric operator. 
If the boundary term in Eq.~\eqref{eq:E-equation-2} is symmetric, then we see from Eq.~\eqref{eq:E-equation-2} that the linearized mode solutions are eigenvectors of a symmetric operator. 
Using the Zerilli-Moncrief function, one can show that the boundary term evaluates to
\begin{align}\label{eq:boundary-term-non-symmetric}
    &-
    \frac{1}{(\omega^2 -\hat{\omega}^2) }
    \bigg.
    \bigg(
    \pmb{\mathscr{R}}\left[\hat{\boldsymbol{y}},\boldsymbol{y}\right] 
    -
    \pmb{\mathscr{R}}\left[\boldsymbol{y}, \hat{\boldsymbol{y}}\right] 
    \bigg)
    =
    \nonumber\\
    &
    \mathscr{s}_0
    \frac{
    \left(\hat{Z} Z'-Z \hat{Z}'\right)}{(\omega^2 - \hat{\omega}^2)}
    +
    \mathscr{s}_1 Z \hat{Z}
    +
    \mathscr{s}_2 \left(Z' \hat{Z} + \hat{Z}' Z \right)
    +
    \mathscr{s}_3 \hat{Z}' Z'
    \,.
\end{align}
The coefficients $\mathscr{s}_i$ depend on the mass and radius of the star, and we present them in Appendix~\ref{appendix:s-coeffs}.
The last three terms in the above equation are symmetric, however, the first term is not symmetric in general.
For isolated stars with QNM boundary conditions, the first term is responsible for outgoing radiation~\cite{1973ApJ...185..685D,FS_stability_rel}.
In this paper, we are interested in the physics of the near zone with  $M \omega \ll 1$ and $R \omega \ll 1$. 
We can use this low-frequency expansion to symmetrize the boundary term in Eq.~\eqref{eq:Z-out-schematic}, using the same logic as in the toy model from Sec.~\ref{sec:frequency-dependent-sturm-liouville-problem}.

We remind the reader that the modes we are interested in obey $Z = Z_{I}$ outside the star. 
Define~\cite{walter1973regular,Fulton1977}
\begin{subequations}\label{eq:Rb-operators}
\begin{align}
    &\boldsymbol{\Delta} \equiv {\rm{det}}
    \begin{vmatrix}
        z_{0,I}(R) & z_{0,I}'(R) \\
        z_{1,I}(R) & z_{1,I}'(R)
    \end{vmatrix}
    \,,\\
     &\boldsymbol{R}_{b,0} \equiv
    Z' z_{1,I}-Z z_{1,I}'  \,,\\
    &\boldsymbol{R}_{b,1} \equiv
    Z z_{0,I}' - Z' z_{0,I}
\end{align}
\end{subequations}
where $z_{i,I}$ are the coefficients of the expansion of $Z$ in Eq.~\eqref{eq:Z-out-schematic}.
By simple algebraic manipulations, one can show that
\begin{subequations}
\begin{align}
    &\boldsymbol{R}_{b,1} = \omega^2 \boldsymbol{R}_{b,0} \,,\\
    &\hat{Z}Z'
    -
    Z\hat{Z}'
    =
    (\omega^2 - \hat{\omega}^2)
    \frac{\boldsymbol{R}_{b,0} \hat{\boldsymbol{R}}_{b,0}}{\boldsymbol{\Delta}}
    \,.
\end{align}
\end{subequations}
Using these identities, we obtain from Eq.~\eqref{eq:boundary-term-non-symmetric}
\begin{align}
    &-
    \frac{1}{(\omega^2 -\hat{\omega}^2) }
    \bigg.
    \bigg(
    \pmb{\mathscr{R}}\left[\hat{\boldsymbol{y}},\boldsymbol{y}\right] 
    -
    \pmb{\mathscr{R}}\left[\boldsymbol{y}, \hat{\boldsymbol{y}}\right] 
    \bigg)
    =
    \mathscr{s}_0
    \frac{\boldsymbol{R}_{b,0} \hat{\boldsymbol{R}}_{b,0}}{\boldsymbol{\Delta}}
    \nonumber\\
    &+
    \mathscr{s}_1 Z \hat{Z}
    +
    \mathscr{s}_2 \left(Z' \hat{Z} + \hat{Z}' Z \right)
    +
    \mathscr{s}_3 \hat{Z}' Z'
    \,.
\end{align}
Substituting this in Eq.~\eqref{eq:E-equation-2}, we see that
\begin{align}\label{eq:orthogonality-E}
    \boldsymbol{E}[\hat{\boldsymbol{y}},\boldsymbol{y}] 
    =
    - (\omega^2 -\hat{\omega}^2) 
    \boldsymbol{I}_0[\hat{\boldsymbol{y}},\boldsymbol{y}] 
    =0
    \,.
\end{align}
If $\omega \neq \hat{\omega}$ then, the eigenfunctions are orthogonal with respect to
\begin{align}
    \boldsymbol{I}_0[\hat{\boldsymbol{y}},\boldsymbol{y}]
    &\equiv 
    \boldsymbol{J}_0[\hat{\boldsymbol{y}},\boldsymbol{y}]
    + 
    \mathscr{s}_0
    \frac{\boldsymbol{R}_{b,0} \hat{\boldsymbol{R}}_{b,0}}{\boldsymbol{\Delta}}
    \nonumber\\
    &+
    \mathscr{s}_1 Z \hat{Z}
    +
    \mathscr{s}_2 \left(Z' \hat{Z} + \hat{Z}' Z \right)
    +
    \mathscr{s}_3 \hat{Z}' Z'
    \,.
\end{align}

This $\boldsymbol{I}_0[\hat{\boldsymbol{y}},\boldsymbol{y}]$ operator is similar to the operator $\boldsymbol{O}[\cdot,\cdot]$ from Eq.~\eqref{eq:O-operator} in the toy-model case.
Just as in the toy-model case, this operator is not positive definite (for $\hat{\boldsymbol{\boldsymbol{y}}}=\boldsymbol{\boldsymbol{y}}$ ), as one can verify explicitly through the expression for $\boldsymbol{O}_{0,\mathrm{DI}}$ in Eq.~\eqref{eq:O0-operator-DI}. 
From this operator, we see that the non-positive definite terms contain products of metric perturbations only, like $\tilde{H}_1 \hat{\tilde{H}}_1$, $K \hat{H}$, and $K \hat{K}$, which are gauge dependent (and thus, hard to interpret physically). 
We note that this non-positive character is also seen in harmonic gauge at 1PN order~\cite{1965ApJ...142.1519C,gittins2025perturbationtheorypostnewtonianneutron}.
Regardless of their interpretation, we mathematically expect generally that operators of this type will not lead to a formally convergent mode sum. 
Nevertheless, the fluid part of the operator in Eq.~\eqref{eq:O0-operator-DI} is positive definite, and, given that most of the tidal dynamics is driven by the fluid sector, we expect that the sum will be ``almost convergent,'' i.e.~any disagreements between direct integration and the mode-sum representation after adding many modes will be small in a relative sense.
Let us then \textit{assume} for now that we can expand solutions in terms of the eigenmodes of the system to see what we find. The consequences of this assumption and non-convergence of the mode  sum will be verified and addressed in Sec.~\ref{sec:results}.

\subsection{Constraints on the tidal field inside the star}\label{sec:constraints-on-tide}
We now turn to the problem of tidal excitation.
Outside the star, we can split the perturbations into a contribution from the star's multipole moment and a term proportional to the tidal field, as in Eq.~\eqref{eq:Z-out-schematic}. 
We now aim to perform an analogous decomposition inside the star. Such a split is necessary only when decomposing the internal perturbations into a mode sum.
If a mode-sum approach is not required, the perturbation equations inside the star can be integrated directly using Eq.~\eqref{eq:Z-out-schematic} as a boundary condition; see~\cite{Pitre:2023xsr,HegadeKR:2024agt,Andersson:2025iyd} for further details.

We denote the perturbations inside the star as 
\begin{align}\label{eq:arbitrary-split-for-tide}
    \boldsymbol{\boldsymbol{y}} &\equiv e^{-i \omega t} Y_{\ell m}\left(H, H_1, H_2, K, V, W\right)
    \nonumber\\
    &=
    \boldsymbol{\boldsymbol{y}}_{\mathrm{T}} + \boldsymbol{\boldsymbol{y}}_{\mathrm{I}}
    \,,
\end{align}
where here we set $H_2 = H$ as required by the linearized Einstein equations, and where $\boldsymbol{\boldsymbol{y}}_{\mathrm{T}}$ is the tidal contribution and $\boldsymbol{\boldsymbol{y}}_{\mathrm{I}}$ is the multipolar contribution.
At this stage, the decomposition described above remains arbitrary. However, we impose four specific conditions on
\begin{align}
    \boldsymbol{\boldsymbol{y}}_{\mathrm{T}} = e^{-i \omega t} Y_{\ell m}\left(H_{T}, H_{1,T},K_{T}, V_{T}, W_{T}\right)
\end{align}
to enable the application of the operator formulation described in Sec.~\ref{sec:action-principle}.
These conditions are:
\begin{enumerate}
    \item [i)] We set $W_{T} = 0$.
    \item [ii)] $V_T$ is obtained through
    \begin{align}
        \Delta \rho_{T} = 0 \implies V_{T} = - \frac{r^2 (H_T - K_T)}{\ell ( \ell + 1)}
        \,.
    \end{align}
    \item [iii)] We demand that $\boldsymbol{\boldsymbol{y}}_{\mathrm{T}}$ satisfies the Hamiltonian and momentum constraints: $\boldsymbol{H}[\boldsymbol{\boldsymbol{y}}_{\mathrm{T}}] =0$, $\boldsymbol{P}_{\sigma}[\boldsymbol{\boldsymbol{y}}_{\mathrm{T}}] = 0$.
    \item [iv)] The functions $(H_{T}, H_{1,T}, K_{T})$ evaluated at the surface of the star should match smoothly to the solution outside the star.
\end{enumerate}

Let us now examine the restrictions imposed on the tidal fields by the Hamiltonian and momentum constraints. In the Regge-Wheeler gauge, the Hamiltonian and momentum constraints are related by the following identity
\begin{align}
    \boldsymbol{E}_{tt}
    &=
    \frac{i e^{\nu -\lambda }}{\omega } \boldsymbol{E}'_{tr}
    +
    \bigg[ 
    -\frac{i e^{\nu -\lambda } \left(r \lambda '-r \nu '-4\right)}{2 r \omega }
    \bigg] \boldsymbol{E}_{tr}
    \nonumber\\
    &-\frac{i \ell (\ell+1) e^{\nu } \boldsymbol{E}_{t \theta} Y_{\ell m}}{r^2 \omega \,\partial_{\theta} Y_{\ell m} } 
    \,.
\end{align}
Therefore, it is sufficient to require that $\boldsymbol{E}_{tr}[\boldsymbol{y}_{T}] = 0$ and $\boldsymbol{E}_{t\theta}[\boldsymbol{y}_{T}] = 0$ in order to satisfy condition (iii).
Define
\begin{align}
    g(r) \equiv 
    \frac{e^{-\lambda } H_{1,T}}{r \omega }+\frac{2 K_{T}}{1+\ell}
    \,,
\end{align}
so that, with this definition, we can solve for $H_{T}$ using $\boldsymbol{E}_{tr}[\boldsymbol{y}_{T}] = 0$ and $\boldsymbol{E}_{t\theta}[\boldsymbol{y}_{T}]=0$ to obtain
\begin{widetext}
\begin{subequations}
\begin{align}
\label{eq:tide-eqn-1}
     &\frac{(1+\ell ) \left(2+e^{\lambda } \ell \right) g}{2 r}
     +
     K_T \bigg[-\frac{\left(-1+e^{\lambda }\right) (1+2 \ell )}{2 r}
     \nonumber\\
     &+\frac{4 \pi  r \left(-[(7+2 \ell ) (e+p)]+e^{\lambda } \left\{e \left(1+8 \pi  r^2 p\right)+p \left[1-\ell  (1+\ell )+8 \pi  r^2 p\right]\right\}\right)}{\ell  (1+\ell )}\bigg]
     \nonumber\\
     &+
     g'
     +
     \frac{\left[(-1+\ell ) \ell -8 \pi  r^2 (e+p)\right] K_T'}{\ell  (1+\ell )}
     =0\,,\\
\label{eq:tide-eqn-2}
     &\frac{1}{4} e^{\lambda } \ell  (1+\ell ) g+H_T+\frac{1}{4} K_T \left[3+2 \ell -e^{\lambda } (1+2 \ell )-8 e^{\lambda } \pi  r^2 p\right]+\frac{r K_T'}{2}
    =0
    \,.
\end{align}
\end{subequations}
\end{widetext}

The above equations involve three unknown functions, $(H_{T}, K_{T}, g)$, so one of these must be specified to obtain a solution. We propose to solve Eqs.~\eqref{eq:tide-eqn-1} and \eqref{eq:tide-eqn-2} using the following strategy:
\begin{itemize}
    \item Equation~\eqref{eq:tide-eqn-2} completely determines $H_{T}$ if we know the solution to $g(r)$ and $K_{T}(r)$.
    \item To solve Eq.~\eqref{eq:tide-eqn-1}, we first choose an arbitrary function $K_{T}(r,c)$ that depends on an arbitrary constant $c$ and matches in a continuous and differentiable manner to the solution for $K$ outside the star. 
    \item With $K_{T}(r,c)$ specified, we integrate Eq.~\eqref{eq:tide-eqn-1} from $r=0$ to a finite radius $r$ to find $g(r,c)$. Regularity at $r=0$ requires that $g(0) = 0$. Hence, this equation can be integrated with the initial condition $g(0) = 0$.
    \item The constant $c$ is obtained by demanding that $g(R,c) = g_{\mathrm{out}}(R)$. This problem can be solved using a simple bisection search for $c$.
\end{itemize}
The sole arbitrary choice in the above formulation is the choice of $K_{T}(r,c)$ inside the star.
This situation is similar to the toy models from Sec.~\ref{sec:toy-models}, where the force inside the domain of the string was not unique.
A simple way to parameterize $K_{T}(r,c)$ inside the star is to set 
\begin{align}\label{eq:Ktide-inside}
    K_{T}(r,c) &= p_0 f_0(r) + p_1 f_1(r) + p_2 f_2(r) +p_3 f_3(r) \nonumber\\
    &+ c f_4(r) \,,
\end{align}
where $f_i(r)$ are any set of linearly independent basis functions that are regular at $r=0$ and combine to yield a power series of the form
\begin{align}
    K_{T}(r,c) = K_T(0,c) + \frac{K_{T}''(0,c)}{2} r^2 + \mathcal{O}(r^4) \,.
\end{align}
For instance, one may choose $f_i(r) = r^{2i}$, following the structure of Taylor expansions near the origin. The constants $p_{i}$ are then determined by requiring that
\begin{align}\label{eq:Ktide-inside-outside-match}
    K_{T}(R,c) &= K_{\mathrm{out}}(R) \,, \nonumber\\ 
    K_{T}'(R,c) &= K_{\mathrm{out}}'(R) \,, \nonumber\\
    K_{T}''(R,c) &= K_{\mathrm{out}}''(R) \, \nonumber\\
    K_{T}'''(R,c) &= K_{\mathrm{out}}'''(R) 
    \,.
\end{align}
The functions $K_{\mathrm{out}}(R),K_{\mathrm{out}}'(R),K_{\mathrm{out}}''(R)$ and $K_{\mathrm{out}}'''(R)$ are provided in the supplementary \texttt{Mathematica} notebook.
\subsection{Mode amplitude evolution}\label{sec:mode-amplitude-evolution}
In this section, we derive an evolution equation for the mode amplitudes.
Let us start with the split provided in Eq.~\eqref{eq:arbitrary-split-for-tide} and understand the evolution of $\boldsymbol{y}_{I}$.
The linearized equations inside the star are given by
\begin{align}
    \boldsymbol{E}_{\alpha \beta} \left[ \boldsymbol{y}_{I} + \boldsymbol{y}_{T} \right] = 0\,, \quad \boldsymbol{E}_{\alpha}\left[ \boldsymbol{y}_{I} + \boldsymbol{y}_{T} \right] = 0\,. 
\end{align}
We rearrange the above equation as
\begin{subequations}\label{eq:tidal-force-eqn-v1}
\begin{align}
    &\boldsymbol{E}_{\alpha \beta}[\boldsymbol{y}_{I}] = - \boldsymbol{E}_{\alpha \beta}[\boldsymbol{y}_{T}] \equiv \boldsymbol{F}_{\alpha \beta}[\boldsymbol{y}_{T}] \,,\\
    &\boldsymbol{E}_{\alpha}\left[ \boldsymbol{y}_{I} \right] = -\boldsymbol{E}_{\alpha}\left[ \boldsymbol{y}_{T} \right] \equiv \boldsymbol{F}_{\alpha}[\boldsymbol{y}_{T}]\,.
\end{align}
\end{subequations}
The tensor $\textbf{F}_{\alpha \beta}= -\boldsymbol{E}_{\alpha \beta} [\boldsymbol{y}_{T}]$ is the ``stress-energy tensor" of the tidal field inside the star, and $\textbf{F}_{\alpha} = -\boldsymbol{E}_{\alpha}\left[ \boldsymbol{y}_{T} \right]$ is the force induced by the tidal field onto the fluid.
If an exact decomposition of the induced field $\boldsymbol{y}_{I}$ and $\boldsymbol{y}_{T}$ were available in general relativity, then tidal fields would satisfy the linearized equations of motion and $\textbf{F}_{\alpha \beta}=  0$. Indeed, this is exactly what happens in Newtonian gravity, where the tidal fields are solutions to the homogeneous Laplace equation~\cite{Poisson-Will}. Such a decomposition does not exist and $\textbf{F}_{\alpha \beta}$ is nonzero. 

To convert the above equations into a forced oscillator problem, we repeat the steps from Sec.~\ref{sec:action-principle} that resulted in the operator formulation [Eq.~\eqref{eq:E-equation-2}].
We recall the operator $\boldsymbol{U}[\hat{\boldsymbol{y}},\boldsymbol{y}_{I}]$ from Eqs.~\eqref{eq:integration-v1}: 
\begin{align}
    \label{eq:integration-v1-for-tide}
    \boldsymbol{U}[\hat{\boldsymbol{y}}, \boldsymbol{y}_{I}] 
    &\equiv
    \int r^2 e^{\lambda/2} dr d \Omega
    e^{\nu(r)/2}
    \bigg[
    \hat{\xi}_{\beta} \boldsymbol{E}^{\beta}[\boldsymbol{y}_{I}] + \frac{\hat{h}_{\alpha \beta} \boldsymbol{E}^{\alpha \beta}[\boldsymbol{y}_{I}] }{16 \pi} 
    \nonumber\\
    &-
    \frac{\hat{\omega} \omega \hat{\mathfrak{h}}_{\rho}}{8\pi} \boldsymbol{P}^{\rho}[\boldsymbol{y}_{I}]
    +
    \frac{\omega^2}{8\pi}
    \hat{\boldsymbol{P}}^{\rho}
    \mathfrak{h}_{\rho}
    \bigg]
    \,.
\end{align}
Substitute Eq.~\eqref{eq:tidal-force-eqn-v1} in Eq.~\eqref{eq:integration-v1-for-tide} and use the fact that solutions satisfy the Hamiltonian and momentum constraints to obtain
\begin{align}\label{eq:tidal-overlap-1}
    \boldsymbol{U}[\hat{\boldsymbol{y}}, \boldsymbol{y}_{I}] \!\!=\!\!
    \int r^2 e^{\frac{\lambda+\nu}{2}} dr d \Omega
    \bigg[
    \hat{\xi}_{\beta} \boldsymbol{F}^{\beta}[\boldsymbol{y}_{T}] + \frac{\hat{h}_{\alpha \beta} \boldsymbol{F}^{\alpha \beta}[\boldsymbol{y}_{T}] }{16 \pi} 
    \bigg]
    \,.
\end{align}
Now, we assume that we can expand $\boldsymbol{y}_{I}$ as a superposition of the modes of the star
\begin{align}\label{eq:mode-sum-approximation-1}
    \boldsymbol{\boldsymbol{y}}_{I} = \sum_{s} \tilde{a}_{s}(\omega) \boldsymbol{\boldsymbol{y}}_{s}(r) e^{- i \omega t} \,,
\end{align}
where $\boldsymbol{\boldsymbol{y}}_{s}$ is the mode solution with frequency $\omega_{s}$.
Substituting Eq.~\eqref{eq:mode-sum-approximation-1} into Eq.~\eqref{eq:tidal-overlap-1} and using Eq.~\eqref{eq:integration-v1} we find
\begin{align}\label{eq:mode-amplitude-evolution}
    &\sum_{s} \tilde{a}_{s}(\omega) \, \boldsymbol{U}
    \left[\hat{\boldsymbol{y}}, \boldsymbol{y}_s e^{-i\omega t}\right]
    \nonumber\\
    &=
    \int r^2 e^{\frac{\lambda+\nu}{2}} dr d \Omega
    \bigg[
    \hat{\xi}_{\beta} \boldsymbol{F}^{\beta}[\boldsymbol{y}_{T}] + \frac{\hat{h}_{\alpha \beta} \boldsymbol{F}^{\alpha \beta}[\boldsymbol{y}_{T}] }{16 \pi} 
    \bigg]
    \,.
\end{align}
Now, we can find the mode amplitude for mode $p$, by setting the arbitrary vector $\hat{y}$ to the eigenvector of the mode $\boldsymbol{y}_p$ to get
\begin{align}\label{eq:mode-amplitude-evolution-p}
    &\sum_{s} \tilde{a}_{s}(\omega) \, \boldsymbol{U}
    \left[\boldsymbol{y}_p, \boldsymbol{y}_s e^{-i \omega t} \right]
    \nonumber\\
    &=
    \int r^2 e^{\frac{\lambda+\nu}{2}} dr d \Omega
    \bigg[
    \xi_{\beta,p} \boldsymbol{F}^{\beta}[\boldsymbol{y}_{T}] + \frac{h_{\alpha \beta,p} \boldsymbol{F}^{\alpha \beta}[\boldsymbol{y}_{T}] }{16 \pi} 
    \bigg]
    \,.
\end{align}
Let us now simplify the left-hand side. Note from Eq.~\eqref{eq:integration-v1} that
\begin{align}\label{eq:eqn-1-sim}
    &\boldsymbol{U}
    \left[\boldsymbol{y}_p, \boldsymbol{y}_s e^{-i \omega t} \right]
    =
    -\omega^2 \int_{r=0}^{R} r^2 e^{\lambda/2} dr\boldsymbol{O}_{0,\DI}[\boldsymbol{y}_p, \boldsymbol{y}_s] 
    \nonumber\\
    &+
    \int_{r=0}^{R} r^2 e^{\lambda/2} dr
    \boldsymbol{O}_{1}[\boldsymbol{y}_p, \boldsymbol{y}_s] 
    +
    \bigg.
    \pmb{\mathscr{R}}\left[\boldsymbol{y}_p, \boldsymbol{y}_s\right] 
    \bigg|_{r=R}
    \,.
\end{align}
Moreover, we also know that the mode eigenfunctions satisfy 
\begin{align}\label{eq:eqn-2-sim}
    &\boldsymbol{U}
    \left[\boldsymbol{y}_s, \boldsymbol{y}_p \right]
    =
    -\omega^2_{p} \int_{r=0}^{R} r^2 e^{\lambda/2} dr\boldsymbol{O}_{0,\DI}[\boldsymbol{y}_s, \boldsymbol{y}_p] 
    \nonumber\\
    &+
    \int_{r=0}^{R} r^2 e^{\lambda/2} dr
    \boldsymbol{O}_{1}[\boldsymbol{y}_s, \boldsymbol{y}_p] 
    +
    \bigg.
    \pmb{\mathscr{R}}\left[\boldsymbol{y}_s, \boldsymbol{y}_p\right] 
    \bigg|_{r=R}
    =
    0\,.
\end{align}
Subtracting Eq.~\eqref{eq:eqn-2-sim} from \eqref{eq:eqn-1-sim}, we see that
\begin{align}
    \boldsymbol{U}
    &\left[\boldsymbol{y}_p, \boldsymbol{y}_s e^{-i \omega t} \right]
    =
    \boldsymbol{U}
    \left[\boldsymbol{y}_p, \boldsymbol{y}_s e^{-i \omega t} \right]
    -
    \boldsymbol{U}
    \left[\boldsymbol{y}_s, \boldsymbol{y}_p \right]
    \nonumber\\
    &=
    \boldsymbol{E}\left[\boldsymbol{y}_p, \boldsymbol{y}_s e^{-i \omega t}\right]
    =
     - (\omega^2-\omega^2_p) 
    \boldsymbol{I}_0[\boldsymbol{y}_p, \boldsymbol{y}_s]
    \,.
\end{align}
Substituting this in Eq.~\eqref{eq:mode-amplitude-evolution-p}, we obtain
\begin{align}
    &\sum_{s} \tilde{a}_{s} (\omega_{p}^2-\omega^2) \boldsymbol{I}_{0}\left[\boldsymbol{y}_{p}, \boldsymbol{y}_{s}\right] 
    \nonumber\\
    &=
    \int r^2 e^{\lambda/2} dr d \Omega
    e^{\nu(r)/2}
    \bigg[
    \xi_{\beta,p} \boldsymbol{F}^{\beta}[\boldsymbol{y}_{T}] + \frac{h_{\alpha \beta,p} \boldsymbol{F}^{\alpha \beta}[\boldsymbol{y}_{T}] }{16 \pi} 
    \bigg]
    \,.
\end{align}
Next, we use the mode normalization condition from Eq.~\eqref{eq:mode-normalization-eqn} to simplify the above equation and obtain an evolution equation for the mode amplitudes
\begin{align}\label{eq:mode-amp-1}
    &\frac{R M^2 \mathscr{N}_{p}}{(R \omega_{p})^2}\tilde{a}_{p} (\omega_{p}^2-\omega^2) \nonumber\\
    &=
    \int r^2 e^{\lambda/2} dr d \Omega
    e^{\nu(r)/2}
    \bigg[
    \xi_{\beta,p} \boldsymbol{F}^{\beta}[\boldsymbol{y}_{T}] + \frac{h_{\alpha \beta,p} \boldsymbol{F}^{\alpha \beta}[\boldsymbol{y}_{T}] }{16 \pi} 
    \bigg]
    \,.
\end{align}
Define the overlap $I_{p}$ so that it satisfies
\begin{align}\label{eq:Is-def}
    &\frac{4 \pi}{2 \ell + 1}d_{\ell m} I_{p}  
    \equiv 
    \frac{1}{M R^{{\ell}}}\int r^2 e^{\lambda/2} dr d \Omega
    e^{\nu(r)/2}
    \bigg[
    \xi_{\beta,p} \boldsymbol{F}^{\beta}[\boldsymbol{y}_{T}] 
    \nonumber\\
    &+ \frac{h_{\alpha \beta,p} \boldsymbol{F}^{\alpha \beta}[\boldsymbol{y}_{T}] }{16 \pi} 
    \bigg] 
    \,.
\end{align}
By substituting the above equation into Eq.~\eqref{eq:mode-amp-1} and simplifying, we obtain
\begin{align}\label{eq:sol-a-freq}
    \tilde{a}_p(\omega) = \frac{4 \pi }{2 \ell + 1} \; \frac{d_{\ell m}}{\mathscr{N}_p \, M/R^{\ell +1}} \, \frac{I_{p}}{1 - (\omega/\omega_p)^2}\,.
\end{align}
The overlap $I_{p}$ determines the solution for the mode amplitude $\tilde{a}_{p}$, just as in Newtonian theory.
Several relativistic versions of the overlap integral have been proposed in literature~\cite{Kuan_2021,Miao_2024}. In Appendix~\ref{appendix:compare-overlap}, we briefly compare Eq.~\eqref{eq:Is-def} we these definitions inspired from Newtonian gravity.
\subsection{Mode-sum representation of the tidal response function}\label{sec:tidal-overlap}
With the mode-amplitudes determined from Eq.~\eqref{eq:sol-a-freq}, we can obtain the tidal response function.
We have tested multiple matching approaches and we see that demanding the continuity of Zerilli-Moncrief function across the surface of the star provides a solution that can accurately satisfy the matching condition.
We note that this prescription is \textit{different} from the one used in Paper I.
Just inside the surface of the star, we can use a mode-sum approximation for $Z$ from Eq.~\eqref{eq:mode-sum-approximation-1}: 
\begin{align}
    Z = \sum_{s} \tilde{a}_{s}(\omega) Z_{s}\,.
\end{align}
Outside the star, we can use Eq.~\eqref{eq:Z-out-schematic} to write
\begin{align}\label{eq:mode-amplitude-match-1}
    &\sum_{s} \tilde{a}_{s}(\omega) Z_{s}(R)
    =   
    Z_{I}
    =
    \frac{4 \pi I_{\ell m}(\omega)}{ (2 \ell +1)}
    \left[ z_{0,I}(R) + z_{1,I}(R) \varepsilon^2  \right]
    .
\end{align}
Using Eq.~\eqref{eq:sol-a-freq} and the definition of the tidal response function,
\begin{align}
    I_{\ell m}(\omega) \equiv 2 K_{\ell m}(\omega) R^{2\ell+1} d_{\ell m}(\omega)\,,
\end{align}
we invert Eq.~\eqref{eq:mode-amplitude-match-1} to obtain
\begin{align}\label{eq:Klm-mode-sum}
    K_{\ell m}(\omega)
    &=
    \frac{2 \pi }{ (2\ell + 1)}
    \sum_{s}
    \frac{I_{s} \mathcal{G}_{s}}{\mathscr{N}_s\left[1 - (\omega/\omega_s)^2\right]}
    \,,
\end{align}
where $I_{s}$ is the overlap defined in Eq.~\eqref{eq:Is-def} and
\begin{align}
    \mathcal{G}_{s} &\equiv 
    \frac{(2 \ell + 1) Z_{s}}{4 \pi M R^{\ell}}
    \bigg[
    z_{0,I}(R) + z_{1,I}(R) (M\omega)^2 \bigg]^{-1}
    \,.
\end{align}
Equation~\eqref{eq:Klm-mode-sum} generalizes the mode-sum approximation used in Newtonian theory and provides a formal approach to derive the harmonic oscillator ansatz routinely used in effective-one-body models~\cite{Steinhoff:2016rfi}.

In deriving the above expressions, we normalized mode solution $y_{s}$ with frequency $\omega_{s}$ via
\begin{align}\label{eq:mode-normalization-eqn}
    \boldsymbol{I}_{0} \left[y_{p}, y_{s}\right] = \frac{R M^2}{(R \omega_{s})^2} \delta_{sp} \mathscr{N}_{s}\,,
\end{align}
to simplify our expressions, where $\delta_{sp}$ is the Kronecker delta, and $\mathscr{N}_{s}$ is an arbitrary normalization factor. The normalization of each mode is arbitrary; only the combination $I_s {\mathcal{G}}_s/{\mathscr{N}}_s$ entering the mode-sum expansion in Eq.~\eqref{eq:Klm-mode-sum} is invariant under a rescaling of the eigenfunctions.

The mode sum of Eq.~\eqref{eq:Klm-mode-sum} is meant to  represent the frequency-dependent response function $K_{\ell m}(\omega)$, but this representation is accurate only if the series is convergent (or, at the very least, asymptotic). As we mentioned before, however, the operator $\boldsymbol{I}_0$, that enters the symmetric operator that defines orthogonality among eigenfunctions is not formally positive definite in general relativity. This implies that the eigenfunctions do not actually form a complete set, and thus, the mode sum might not converge, as we will check in the next section. We emphasize, however, that our statements about non-convergence refer specifically to the mode-sum construction developed here in Regge--Wheeler gauge, with near-zone boundary conditions truncated at ${\cal{O}}(\omega^2)$. In this setting, the dominant obstruction is the non-positive-definite character of the orthogonality operator, while the low-frequency truncation further degrades the accuracy of the higher-frequency modes retained in the sum.

\subsection{Summary and numerical implementation}\label{sec:summary-and-num}

The previous subsections were rather heavy mathematically, so we feel it is useful to provide a summary here to facilitate the numerical implementation of the relativistic tidal response calculation. Our implementation summary is as follows:
\begin{itemize}
    \item \textbf{Mode solutions:} Determine the mode frequencies and profiles for the metric and fluid perturbations by integrating the master equations [Eq.~\eqref{eq:master-equation-matrix}] with the boundary condition specified in Eq.~\eqref{eq:Z-out-schematic}. Normalize the mode solutions using Eq.~\eqref{eq:mode-normalization-eqn}. 

    \item \textbf{Tidal field inside the star:} Specify a profile for $K_{T}(r,c)$ within the star [Eq.~\eqref{eq:Ktide-inside}], and determine the coefficients $p_i$ using Eq.~\eqref{eq:Ktide-inside-outside-match}. Numerically solve for $g(r)$ by integrating Eq.~\eqref{eq:tide-eqn-1} with the initial condition $g(0) = 0$. The constant $c$ is obtained by requiring $g(R) = g_{\mathrm{out}}(R)$. Compute $H_{T}(r)$ inside the star using Eq.~\eqref{eq:tide-eqn-2}.

    \item \textbf{Overlap integral:} Employ the profiles of the mode solutions and the tidal field inside the star to calculate $I_{s}$ [Eq.~\eqref{eq:Is-def}].

    \item \textbf{Tidal response:} Apply Eq.~\eqref{eq:Klm-mode-sum} to obtain the mode-sum approximation of the tidal response function.
\end{itemize}
The exterior solutions for all the metric perturbations are provided in the supplementary \texttt{Mathematica} notebook.
\section{Numerical Results and Convergence of the Mode sum}\label{sec:results}
In this section, we implement the mode-sum approach discussed in Sec.~\ref{sec:summary-and-num} for different nuclear EoSs. 
In Sec.~\ref{sec:compare-mode-QNM}, we compare the $f$-mode frequencies and solutions with  QNM solutions. By QNM solutions, we here mean the complex frequencies obtained by using outgoing wave conditions at null-infinity.
Next, we examine how the tidal field within the star varies as a function of compactness in Sec.~\ref{sec:tide-inside-the-star}.
We then analyze the convergence of the mode sum relative to the dynamical tidal response obtained through direct numerical integration in Sec.~\ref{sec:convergence-of-mode-sum}.
 
\subsection{Comparing mode solutions to QNMs}\label{sec:compare-mode-QNM}

\begin{figure}[tb]
    \centering
    \includegraphics[width=0.99\linewidth]{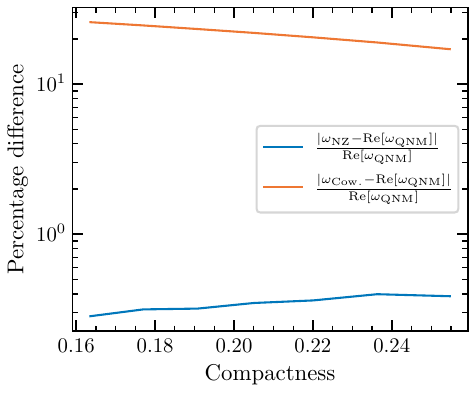}
    \caption{Comparison between the $f$-mode frequency obtained using the near-zone condition of this paper (blue), the frequency obtained using the Cowling approximation (orange), and the real part of the $f$-mode QNM frequency. Observe that the matched frequencies are highly accurate relative to the QNM frequencies, while the Cowling approximation presents the usual ${\cal{O}}(10 \%)$ discrepancies.}
    \label{fig:mode-frequency-compare}
\end{figure}
\begin{figure*}[thp!]
    \centering
    \includegraphics[width=1.98\columnwidth]{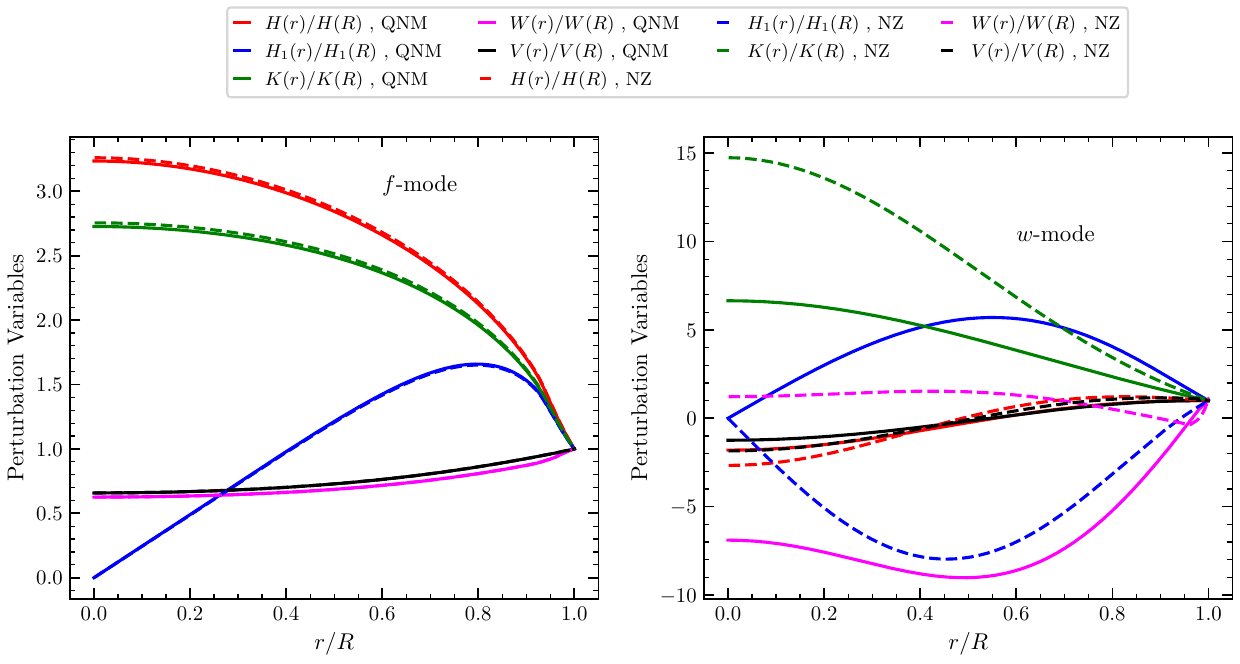}
    \caption{
    Comparison between profiles of various metric and fluid variables for the $f$-mode eigenfunctions (left panel) and the $w$-mode eigenfunctions (right panel) obtained using the near-zone boundary condition (dashed lines) with the real part of the QNM eigenfunctions for a star of mass $M=1.4 M_{\odot}$. 
    Observe that the near-zone boundary condition can accurately capture the behavior of the $f$-mode QNM but cannot reproduce the high-frequency $w$ mode.
    }
    \label{fig:eigenfunction_compare}
\end{figure*}

Figure~\ref{fig:mode-frequency-compare} presents a comparison between the $f$-mode frequencies obtained using the procedure described in Sec.~\ref{sec:tidal-overlap} and the real part of the QNM frequencies for the SLY EoS as a function of compactness.
The orange curve in the figure represents the mode frequencies calculated using the relativistic Cowling approximation, which is well-known to exhibit errors of $\mathcal{O}(10-25) \%$ when compared to the QNM frequencies.
We remind the reader that, in the Cowling approximation, the metric perturbations inside the star are ``frozen,'' and one evolves only the fluid variables~\cite{Kunjipurayil_2022}.
The blue curve demonstrates that the mode frequencies calculated using the near-zone approximation closely match the real part of the QNM frequency, with errors smaller than $1$\%~\cite{Lindblom_1997,Andersson:2025iyd}.

\begin{figure*}[thp!]
    \centering
    \includegraphics[width=0.99\columnwidth]{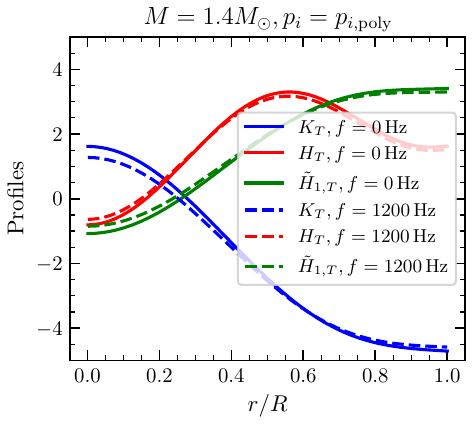}
    \includegraphics[width=0.99\columnwidth]{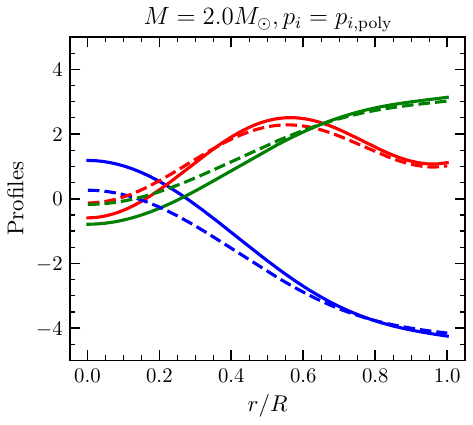}
    \includegraphics[width=0.99\columnwidth]{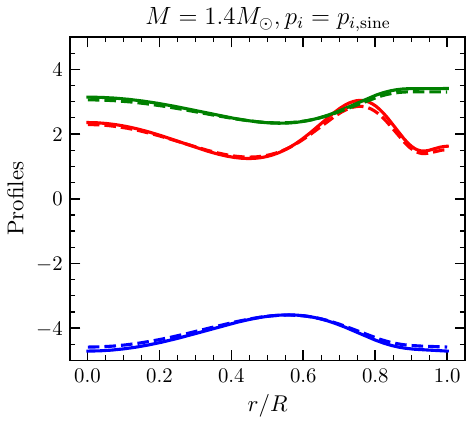}
    \includegraphics[width=0.99\columnwidth]{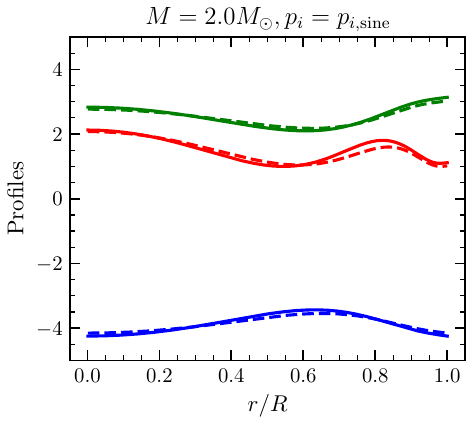}
    \caption{
    Variation in the profile of the tidal field inside the star for different choices of the basis functions [Eq.~\eqref{eq:basis-choice}]. The top (bottom) panels compare the profiles with the polynomial (sine) basis for stars of mass $1.4 M_{\odot}$ (left) and $2 M_{\odot}$ (right) for two different frequencies (bold and dashed curves). Observe that the compactness of the star and the choice of the basis functions impact the profiles, which is not problematic as these quantities are gauge dependent.}
    \label{fig:tides-inside-the-star}
\end{figure*}
In the left panel of Fig.~\ref{fig:eigenfunction_compare}, we compare the profiles of the different metric and fluid variables (dashed lines) inside the star with the real part of the profiles obtained using the QNM boundary conditions (solid lines) for a star of mass $M=1.4 M_{\odot}$. Observe that the mode eigenfunctions, obtained with the near-zone boundary condition, closely match the QNM eigenfunctions.
The results of Fig.~\ref{fig:mode-frequency-compare} and the left panel of Fig.~\ref{fig:eigenfunction_compare} demonstrate that the near-zone boundary condition can accurately capture the behavior of the $f$ mode for compact relativistic stars~\cite{Lindblom_1997,Andersson:2025iyd}.

In addition to the $f$-mode, the near-zone boundary condition also admits $w$-mode-like and $p$-mode-like solutions inside the star. In the right panel of Fig.~\ref{fig:eigenfunction_compare}, we compare the first such solution with the real part of the corresponding QNM, where the latter is computed using the mixed numerical/WKB method of Ref.~\cite{Kokkotas:1992}. Unlike the $f$-mode, the near-zone $w$-like solution does not closely track the QNM profile. This is expected: the near-zone boundary condition is a low-frequency approximation, whereas the $w$-modes are high-frequency, strongly-damped spacetime modes. This panel, therefore, clearly describes the limitations of the near-zone matching method. 
\subsection{Tidal field inside the star}\label{sec:tide-inside-the-star}

In this section, we compare how different choices of the function $K_{T}(r,c)$ [Eq.~\eqref{eq:Ktide-inside}] impact the profiles for the tidal field inside the star.
For clarity, we examine two distinct choices for the basis function coefficients $p_{i}$
\begin{subequations}\label{eq:basis-choice}
\begin{align}
    &p_{i,\mathrm{poly}} = \left(\frac{r}{R}\right)^{2i} \,,\\
    &p_{i,\mathrm{sine}}  =
    \begin{cases}
        1, i=0 \,,\\
        \sin \left[\pi \left(\frac{r}{R}\right)^{2i}\right], i> 0 \,.
    \end{cases}
\end{align}
\end{subequations}
The polynomial basis function is motivated by a PN expansion of the tidal field inside the star. We can interpret each polynomial basis function as a PN counting parameter,
\begin{align}
    p_{i,\mathrm{poly}} = \left(\frac{r}{R}\right)^{2i} \propto \left( \frac{4\pi}{3} \frac{e_c r^3}{r} \right)^{i}
    \propto 
    \left( \frac{M_c}{r} \right)^{i}
\end{align}
where $e_c$ is the central energy density and $M_c = \frac{4\pi}{3} e_c r^3$ is an approximate measure of the mass inside radius $r$. The sine basis function is chosen simply to explore results with a basis function that is clearly very different from a polynomial. 

Figure~\ref{fig:tides-inside-the-star} illustrates how the different choices for the basis functions affect the tidal field inside the star.
Each panel presents the profiles of the metric functions $(H_{T},K_{T},H_{1,T})$ inside the star for two frequencies: $\omega = 0\,\mathrm{Hz}$ (solid lines) and $\omega = (2\pi) \times 1200\, \mathrm{Hz}$ (dashed lines).
These frequency choices show how the tidal field inside the star changes during the early and very late inspiral, respectively.
The top (bottom) row shows the profiles for the $p_{i} = p_{i,\mathrm{poly}}$ ($p_{i} = p_{i,\mathrm{sine}})$ choice, with $M = 1.4 M_{\odot}$ in the top (bottom) left panel and $M = 2.0 M_{\odot}$ in the top (bottom) right panel.
In each panel, the frequency does not significantly affect the profiles of the metric functions.
The figure demonstrates that, for both compactnesses, the choice of basis functions substantially influences the profile of the tidal field inside the star. Such a result is not a problem, when one realizes that the interior profile of these metric functions is not a direct observable, as it is clearly gauge dependent.
This change in the tidal field as a function of compactness is not accounted for in studies that use a simple Newtonian prescription for the tidal field inside the star~\cite{Gao:2025aqo}. 
\subsection{Convergence of the mode sum}\label{sec:convergence-of-mode-sum}

\begin{figure}[thb]
    \centering
    \includegraphics[width=0.99\linewidth]{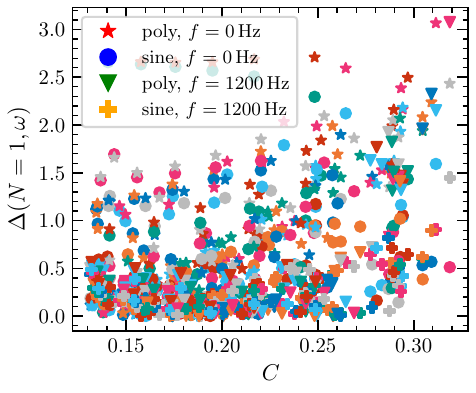}
    \caption{Difference between the $f$-mode contribution to the mode-sum approximation and the direct matching calculation for different EoSs.
    The different markers in the plot represent different choices for the tidal basis functions and the driving frequency $f = \omega/2\pi$.}
    \label{fig:k2-f-mode-difference}
\end{figure}
The next test of our formalism is to understand how well the static limit ($\omega = 0$) of the mode sum [Eq.~\eqref{eq:Klm-mode-sum}] converges to the tidal response function obtained through a direct matching calculation. For simplicity, we assume that the EoS is piecewise polytropic, and hence, the mode solutions do not contain $g$-mode contributions. We expect our results to be robust to other choices of EoS without non-trivial structure in the speed of sound.  
To assess our convergence, we use a diagnostic similar to Eq.~\eqref{eq:toy-mode-convergence-diagnostic}:
\begin{align}
    \Delta_{\mathrm{mode}}(N,\omega) &\equiv \left|1 - \frac{2\pi}{5 K_{22}(\omega)}\sum_{s=1}^{N}
    \frac{I_{s} \mathcal{G}_{s}}{\mathscr{N}_s\left[1 - (\omega/\omega_s)^2\right]}
    \right|
    \nonumber\\
    &\times 100
    \,,
\end{align}
where the function $K_{22}(\omega)$ is calculated by using the direct matching calculation from~\cite{HegadeKR:2024agt}.
Physically, we expect that the dominant contribution to the mode sum arises from the $f$-mode contribution.
In Fig.~\ref{fig:k2-f-mode-difference} we show how $\Delta_{\mathrm{mode}}(N=1,\omega)$ varies for different piecewise polytropic EoSs\footnote{Namely, ALF2, AP3, BSK20–26, BSP, DD2, DD2Y, DDHd, DDME2, DDME2Y, ENG, FSUGarnet, G3, IOPB, MPA1, Model1, Rs, SINPA, SK255, SK272, SLY2, SLY230A, SLY4, SLY9, SLy, SkI4, SkI6, SkMP, Ska, Skb~\cite{Read_2009}.} (colored markers), tidal basis choices (polynomial and sinusoidal basis from Eq.~\eqref{eq:basis-choice}) and the driving frequency $f=\omega/2\pi$. Observe that the relative fractional difference is less than $\sim 3\%$ for different EoSs across a wide range of compactnesses.
This suggests that the $f$-mode contribution is robust and closely reproduces the matching calculation.

\begin{figure}[t]
    \centering
    \includegraphics[width=0.99\linewidth]{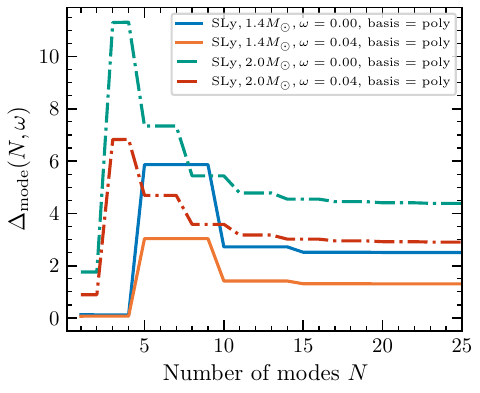}
    \caption{Non-convergence of the relativistic mode sum. As we keep a large number of modes in the mode sum, the mode sum does not converge to the direct matching calculation ($\Delta_{\mathrm{mode}} \neq 0$). This behavior is reminiscent of the non-convergence we observed in the toy model, see the orange curve in the right panel of Fig.~\ref{fig:toy-model-plus-minus}.}
    \label{fig:k2-non-convergence}
\end{figure}

Now, we test the convergence of the mode sum as we add modes with frequencies higher than the $f$-mode frequency. In Fig.~\ref{fig:k2-non-convergence}, we plot $\Delta_{\mathrm{mode}}(N,\omega)$ for the SLy EoS and different masses and frequencies with the polynomial tidal basis. 
As we see from the figure, $\Delta_{\mathrm{mode}}(N,\omega) \neq 0$, which implies that the mode sum does not seem to converge to the direct matching calculation.
Moreover, the behavior observed in the figure is very similar to the non-convergence observed in the $\epsilon = -0.1$ case of the toy model (compare to the right panel of Fig.~\ref{fig:toy-model-plus-minus}).

As discussed above, this behavior is likely due to the non-positive-definiteness of the operators in the strong-field regime. As we argued before, the operators are ``almost'' positive definite, in the sense that the terms that break this condition contribute little to the operator. Thus, we may expect that the mode sum should ``almost'' converge, but not quite. This is exactly what we see in Fig.~\ref{fig:k2-non-convergence}. 

One may worry that the non-convergence we find is due to numerical error that may accumulate in our calculation. For example, while the modes are theoretically orthogonal with respect to $\boldsymbol{I}_0$ 
, in practice we achieve orthogonality to $\mathcal{O}(10^{-5}-10^{-2})$ depending on the mode number. The orthogonality is also worse for higher values of $N$, where there are a large number of oscillations. 
We have however explored this numerical source of error, reduced it by a more careful numerical treatment, and yet we find that the non-convergence remains and Fig.~\ref{fig:k2-non-convergence} is unaffected. We thus conclude that numerics are not the root of the problem. 

Figures~\ref{fig:k2-f-mode-difference} and~\ref{fig:k2-non-convergence} show that the value of the relativistic mode-sum does not lie in systematic convergence with increasing mode number. Instead, its value lies in the fact that the dominant $f$-mode already captures the response to good accuracy. Once higher-frequency modes are included, the sum need not approach the direct matching calculation, because the orthogonality operator is not positive definite and the low-frequency boundary expansion becomes progressively less accurate. The loss of convergence should therefore not be interpreted as a purely numerical artifact. The toy model already showed that non-convergence is expected once positivity is lost; numerical loss of orthogonality at high mode number likely worsens the problem, but does not create it. Regardless of these issues, from an observational perspective, the accuracy of the mode-sum approach when one retains only the $f$-mode contribution should be more than adequate.

\section{Conclusions}\label{sec:conclusions}

In this paper, we implemented the relativistic mode-sum framework of~\cite{HegadeKR:2025qwj} for non-rotating neutron stars in Regge--Wheeler gauge and used it to test the strong-field fate of the Newtonian oscillator picture. The answer is useful, but not na\"ive. Near-zone boundary conditions accurately capture the dominant $f$ mode~\cite{Lindblom_1997,Andersson:2025iyd}, and the corresponding $f$-mode contribution reproduces the direct matching calculation of~\cite{HegadeKR:2024agt} at the few-percent level across the equations of state studied here.

At the same time, the strong-field norm that governs mode orthogonality is not positive definite, and the full mode-sum is therefore not expected to converge strictly to the direct matching result. More precisely, this statement refers to the particular mode-sum construction developed here in Regge--Wheeler gauge, with near-zone boundary conditions truncated at $O(\omega^2)$. In this formulation, the dominant obstruction is the non-positive-definite character of the norm operator, while the low-frequency truncation further limits the accuracy of the higher-frequency modes. This non-positivity stands in stark contrast to the positive-definite norm in the Newtonian problem. Our results therefore show that, while mode decomposition is a powerful organizing principle in Newtonian theory, it must be used with considerably more care for relativistic stars.

Our analysis also shows that relativistic overlap integrals and interior tidal fields can be defined systematically, but that both inherit a genuine ambiguity associated with how the tidal field is extended into the stellar interior. The practical lesson is that, in this setting, the relativistic mode sum is most useful as a low-mode approximation, not as a convergent expansion over arbitrarily many modes.

Because the aforementioned obstruction arises in the specific mode-sum construction studied here, several interesting possibilities remain open. One is that a different gauge, or a formulation based on different master variables, may admit a better-behaved relativistic mode expansion. Another is that the exact gauge-invariant gravitational-wave observable may not be representable as a purely discrete mode sum, but instead may require a discrete modal contribution supplemented by a genuinely non-modal remainder. A third, and perhaps more practical, possibility is that the physically-relevant observable is already captured accurately enough (for observational purposes) by a controlled low-mode truncation, even in the absence of strict convergence of the full expansion. Our results for the dominant $f$ mode strongly support this last interpretation, while the first two remain important open theoretical questions.

More broadly, these results clarify how mode-based descriptions of relativistic dynamical tides should be used in waveform modeling. The strength of the approach lies in its ability to capture the physically relevant low-lying fluid dynamics, especially the dominant $f$ mode, rather than in any formal convergence of the full strong-field expansion. Our results for the $f$ mode, combined with recent progress on the $g$-mode sector~\cite{Andersson:2025iyd}, provide a practical route to characterizing the tidal response of quasi-circular binary neutron star systems. A concrete path toward relativistic mode-based tidal models now exists, while making equally clear where Newtonian intuition fails.
\acknowledgements
We thank Nils Andersson for illuminating discussions.
T.V. and K.J.K. acknowledge support from NSF grants 2012086 and 2309360, the Alfred P. Sloan Foundation through grant number FG-2023-20470, and the BSF through award number 2022136.
H.Y. is supported by NSF award No. PHY-2308415, CAREER award No. PHY-254157, and Montana NASA EPSCoR Research Infrastructure Development under award No. 80NSSC22M0042.
N.Y.~acknowledges support from the Simons Foundation through Award No. 896696, the Simons Foundation International through Award No. SFI-MPS-BH-00012593-01, the NSF through Grant No.
PHY-25-12423.
Large language model Gemini was used to create the cartoon in Fig~\ref{fig:toy-problem-cartoon}.
\appendix
\section{Mode amplitudes for the toy model}\label{appendix:mode-amplitudes-toy-model}

Below we present the mode amplitudes for the toy problem using the two external functions of Eq.~\eqref{eq:F-choices-toy-model}:
\begin{widetext}
\begin{subequations}
\begin{align}
    \label{eq:ak-poly}
    &a_{k,\mathrm{poly}} = 
    \frac{1}{(\omega^2 - \omega_k^2)}
    \bigg[
    \frac{f_{\text{ext}} \left(32 \omega _k^2-32 \omega ^2\right)+f^{\mathrm{ext}}_1 \left(\omega ^2 \left(16 \pi -4 \pi ^2 (-1)^k \omega _k\right)-16 \pi  \omega _k^2\right)}{ ( \pi \omega_k )^3}
    \bigg]
    \,,\\
    \label{eq:ak-sine}
    &a_{k,\mathrm{sine}} =
    -\frac{4 \left(18 f_{\text{ext}} \left(4 k^2+4 k-\omega ^2+1\right)+(-1)^k (2 k+1) f_{1}^{\text{ext}} \left(\omega ^2-36\right)\right)}{\pi  (2 k+1) \left(4 k^2+4 k-35\right) \left(\omega ^2-(2 k+1)^2\right)}
    \,.
\end{align}
\end{subequations}
\end{widetext}
These mode amplitudes are \textit{exact}. 
\section{Linearized equations of motion and metric reconstruction}\label{appendix:DI-eom}
The linearized equations of motion inside the star are~\cite{1985ApJ...292...12D,Kunjipurayil_2022} 
\begin{widetext}
\begin{subequations}
\begin{align}
    &r \tilde{H}_1' = 
    2 e^{\lambda } H-e^{\lambda } K+\tilde{H}_1 \left(-1-\ell -2 e^{\lambda } b+4 e^{\lambda } \pi  r^2 e-4 e^{\lambda } \pi  r^2 p\right)+16 e^{\lambda } \pi  (e+p) V
    \,,\\
    &r K' = -2 H-\frac{1}{2} \ell  (1+\ell ) \tilde{H}_1+\frac{1}{2} K \left(-3+e^{\lambda }-2 \ell +8 e^{\lambda } \pi  r^2 p\right)-8 e^{\lambda /2} \pi  (e+p) W
    \,,\\
    &r W' = -e^{\lambda /2} r^2 H+e^{\lambda /2} r^2 K-e^{\lambda /2} \ell  (1+\ell ) V+(-1-\ell ) W-\frac{e^{\frac{\lambda -\nu }{2}} r^2 (e+p) X}{p \Gamma }
    \,,\\
    &r X' =
    \frac{1}{4} e^{-\frac{\nu }{2}} \left(e^{\nu } \ell  (1+\ell )+2 r^2 \omega ^2\right) \tilde{H}_1
    +\frac{1}{4} e^{\nu /2} \text{K} \left(5-3 e^{\lambda } \left(1+8 \pi  r^2 p\right)\right)
    -\frac{1}{2} e^{\nu /2} \text{H} \left(-3+e^{\lambda } \left(1+8 \pi  r^2 p\right)\right)
    \nonumber\\
    &+
    \frac{e^{\nu /2} \ell  (1+\ell ) \left(-1+e^{\lambda } \left(1+8 \pi  r^2 p\right)\right) V}{2 r^2}+X \left(-\ell -\frac{r \left(1+ de/dp\right) p'}{e+p}\right)
    \nonumber\\
    &+W \left(\frac{e^{\frac{1}{2} (-\lambda -\nu )} \left(-3 e^{\nu }+e^{\lambda } \left(2 r^2 \omega ^2+e^{\nu } \left(3+8 \pi  r^2 (e+2 p)\right)\right)\right)}{2 r^2}-4 e^{\frac{\lambda +\nu }{2}} \pi  r p'-\frac{e^{\frac{1}{2} (-\lambda +\nu )} \left(1+e^{\lambda } \left(1+8 \pi  r^2 p\right)\right) \lambda '}{4 r}\right)
    \,,
\end{align}
\end{subequations}
where 
\begin{subequations}
\begin{align}
    Q &\equiv \frac{1}{2} e^{-\lambda } \left(-1+e^{\lambda }+8 e^{\lambda } \pi  r^2 p\right)
    \,,\\
    b&\equiv \frac{1}{2} e^{-\lambda } \left(-1+e^{\lambda }\right) \,.
\end{align}
\end{subequations}
One can reconstruct $H$ and $V$ from $(\tilde{H}_1, K, W, X)$ 
by using the following identities
\begin{subequations}
\begin{align}
    H &=
    \frac{e^{-\nu } \tilde{H}_1 \left(-e^{\nu } \left(-1+e^{\lambda }\right) \ell  (1+\ell )+4 r^2 \omega ^2-8 e^{\lambda +\nu } \ell  (1+\ell ) \pi  r^2 p\right)}{4 \left(-3+e^{\lambda } \left(1+\ell +\ell ^2+8 \pi  r^2 p\right)\right)}
    \nonumber\\
    &+\frac{e^{-\nu } K \left(3 e^{\nu }+e^{2 \lambda +\nu } \left(1+8 \pi  r^2 p\right)^2+2 e^{\lambda } \left(2 r^2 \omega ^2-e^{\nu } \left(\ell +\ell ^2+16 \pi  r^2 p\right)\right)\right)}{4 \left(-3+e^{\lambda } \left(1+\ell +\ell ^2+8 \pi  r^2 p\right)\right)}+\frac{8 e^{\lambda -\frac{\nu }{2}} \pi  r^2 (e+p) X}{-3+e^{\lambda } \left(1+\ell +\ell ^2+8 \pi  r^2 p\right)}
    \,,\\
    V &= 
    -\frac{\tilde{H}_1 \left(e^{\nu } \left(-1+e^{\lambda }\right) \ell  (1+\ell )-4 r^2 \omega ^2+8 e^{\lambda +\nu } \ell  (1+\ell ) \pi  r^2 p\right)}{4 \omega ^2 \left(-3+e^{\lambda } \left(1+\ell +\ell ^2+8 \pi  r^2 p\right)\right)}
    \nonumber\\
    &+\frac{K \left(3 e^{\nu }+e^{2 \lambda +\nu } \left(1+8 \pi  r^2 p\right)^2+2 e^{\lambda } \left(2 r^2 \omega ^2-e^{\nu } \left(\ell +\ell ^2+16 \pi  r^2 p\right)\right)\right)}{4 \omega ^2 \left(-3+e^{\lambda } \left(1+\ell +\ell ^2+8 \pi  r^2 p\right)\right)}+\frac{e^{-\frac{\lambda }{2}+\nu } \left(1-e^{\lambda } \left(1+8 \pi  r^2 p\right)\right) W}{2 r^2 \omega ^2}
    \nonumber\\
    &+\frac{e^{\nu /2} \left(3-e^{\lambda } \left(1+\ell +\ell ^2-8 \pi  r^2 e\right)\right) X}{\omega ^2 \left(-3+e^{\lambda } \left(1+\ell +\ell ^2+8 \pi  r^2 p\right)\right)}
    \,,
\end{align}
\end{subequations}
and the linearized Einstein equations also require $H_2 = H$. 

Outside the star, we can reconstruct all the metric variables from $Z$ 
\begin{align} 
    &K = \frac{r^{-2-\ell } \left(24 M^2+6 \left(-2+\ell +\ell ^2\right) M r+\ell  \left(-2-\ell +2 \ell ^2+\ell ^3\right) r^2\right) Z}{2 \left(6 M+\left(-2+\ell +\ell ^2\right) r\right)}+r^{-1-\ell } (-2 M+r) Z'
    \,,\\
    &\tilde{H}_{1} =
    r^{-\ell } \left(\frac{\left(-6 M^2-3 \left(-2+\ell +\ell ^2\right) M r+\left(-2+\ell +\ell ^2\right) r^2\right) Z}{(2 M-r) r \left(6 M+\left(-2+\ell +\ell ^2\right) r\right)}-Z'\right)
    \,,\\
    &H = 
    \left(\frac{r^{-2-\ell } \left(-72 M^3-36 \left(-2+\ell +\ell ^2\right) M^2 r-6 \left(-2+\ell +\ell ^2\right)^2 M r^2-\ell  (1+\ell ) \left(-2+\ell +\ell ^2\right)^2 r^3\right)}{4 \left(6 M+\left(-2+\ell +\ell ^2\right) r\right)^2}-\frac{r^{2-\ell } \omega ^2}{4 M-2 r}\right) Z
    \nonumber\\
    &+\frac{r^{-1-\ell } \left(6 M^2+3 \left(-2+\ell +\ell ^2\right) M r-\left(-2+\ell +\ell ^2\right) r^2\right) Z'}{2 \left(6 M+\left(-2+\ell +\ell ^2\right) r\right)}
    \,.
\end{align}
\end{widetext}
\section{Operators}\label{appendix:operators-proof}
The expressions for the operators appearing in Eqs.~\eqref{eq:operator-form-hamiltonian-general-gauge}-\eqref{eq:Lagrangian-integrated-by-parts-1} are provided below
\begin{widetext}
\begin{dgroup}\label{eq:operators-for-equation}
\begin{dmath}
    \pmb{\mathcal{A}}_{H}[\hat{\boldsymbol{y}},\boldsymbol{y}] \hiderel{=}
    \frac{e^{-\lambda + \nu/2} r^{2\ell -2}}{8 \pi}
    \bigg[
    8 e^{\lambda } \ell  (1+\ell ) \pi  \hat{H} (e+p) V-\frac{4 e^{\lambda /2} \pi  \hat{H} (e+p) W \left(-1+e^{\lambda }+8 e^{\lambda } \pi  r^2 p-2 (1+\ell ) c_s^2\right)}{c_s^2}+H \left(\ell  \left(-2+e^{\lambda } (1+\ell )\right) \hat{H}-2 r \hat{H}'\right)+K \left(-\frac{1}{2} \hat{H} \left(\ell  \left(1+2 \ell +e^{\lambda } (2+\ell )\right)+8 e^{\lambda } \pi  r^2 e+8 e^{\lambda } (3+\ell ) \pi  r^2 p\right)-(1+\ell ) r \hat{H}'\right)-\frac{1}{2} r \left(\hat{H} \left(-1+e^{\lambda }+2 \ell +8 e^{\lambda } \pi  r^2 p\right)+2 r \hat{H}'\right) K'+8 e^{\lambda /2} \pi  r \hat{H} (e+p) W'
    \bigg]
    \,,
\end{dmath}
\begin{dmath}
    \pmb{\mathcal{A}}_{P}[\hat{\boldsymbol{y}},\boldsymbol{y}] \hiderel{=} 
    \frac{r^{2\ell - 2} e^{-\lambda - \nu/2}}{16 \pi }
    \Bigg[
    4 r^2 H \hat{\tilde{H}}_1+16 e^{\lambda /2} \pi  r^2 \hat{\tilde{H}}_1 (e+p) W-e^{-\lambda } r^3 \hat{\tilde{H}}_1 \tilde{H}_1'+r^2 K \left(e^{\lambda } \hat{\tilde{H}}_1 \left(-1+4 \pi  r^2 e-4 \pi  r^2 p\right)-r \hat{\tilde{H}}_1'\right)+e^{-\lambda } r^2 \tilde{H}_1 \left(\hat{\tilde{H}}_1 \left(-2-e^{\lambda }-2 \ell +e^{\lambda } \ell +e^{\lambda } \ell ^2+12 e^{\lambda } \pi  r^2 e+4 e^{\lambda } \pi  r^2 p\right)-r \hat{\tilde{H}}_1'\right)+r^3 \hat{\tilde{H}}_1 K'
    \Bigg]
    \,,
\end{dmath}
\begin{dmath}
    \pmb{\mathcal{R}}_{H}[\hat{\boldsymbol{y}},\boldsymbol{y}] \hiderel{=}
    \frac{e^{-\lambda + \nu/2} r^{2\ell}}{8 \pi}
    \Bigg[
    e^{\lambda /2} r \hat{H} \left(2 H+(1+\ell ) K+r K'\right)
    \Bigg]
    \,,
\end{dmath}
\begin{dmath}
    \pmb{\mathcal{R}}_{P}[\hat{\boldsymbol{y}},\boldsymbol{y}] \hiderel{=}
    \frac{r^{2\ell} e^{-\lambda - \nu/2}}{16 \pi }
    \Bigg[ 
    e^{-\frac{\lambda }{2}} r^3 \hat{\tilde{H}}_1 \left(\tilde{H}_1+e^{\lambda } K\right)
    \Bigg]
    \,,
\end{dmath}
\begin{dmath}\label{eq:O0-operator-DI}
    \boldsymbol{O}_{0,\DI}[\hat{\boldsymbol{y}},\boldsymbol{y}] \hiderel{=}
    e^{-\nu/2} r^{2\ell-2}
    \bigg[ 
    -\frac{e^{-\lambda } \ell  (1+\ell ) r^2 \tilde{H}_1 \hat{\tilde{H}}_1}{16 \pi }+\frac{r^2 K (2 \hat{H}-\hat{K})}{16 \pi }+\frac{r^2 H \hat{K}}{8 \pi }+\ell  (1+\ell ) (e+p) V \hat{V}
    \nonumber\\
    +(e+p) W \hat{W}
    \bigg]
    \,,
\end{dmath}
\begin{dmath}
    \pmb{\mathcal{R}}[\hat{\boldsymbol{y}},\boldsymbol{y}] \hiderel{=}
    r^{2 \ell+1} e^{\frac{1}{2} (-\nu -\lambda)}
    \bigg[ 
    \frac{r^2 \omega ^2 \tilde{H}_1 \hat{K}}{8 \pi }+\frac{e^{\nu } (1+\ell ) K \hat{K}}{16 \pi }+\frac{e^{\nu } r \hat{K} H'}{8 \pi }+\frac{e^{\nu } r \hat{K} K'}{16 \pi }-\frac{e^{\nu } H \left(2 \hat{H}+\hat{K}-e^{\lambda } \hat{K}+r \hat{K}'\right)}{8 \pi }
    \bigg]
    +
    \delta \pmb{\mathcal{R}}
    \,,
\end{dmath}
\end{dgroup}
where $\delta \pmb{\mathcal{R}}$ contains terms that evaluate to zero at $r=R$ which are not needed in this paper. 
The expression for $\boldsymbol{O}_1$ is too long and complicated to be displayed here.
We include it in the supplementary \texttt{Mathematica} notebook.
\section{Expressions for $\mathscr{s}_i$}\label{appendix:s-coeffs}
The coefficients are
\begin{subequations}
\begin{align}
    \mathscr{s}_0 &=
    \frac{(1-2 C) \ell  \left(-2-\ell +2 \ell ^2+\ell ^3\right)}{64 \pi }
    \,,\\
    \mathscr{s}_1 &= 
    \frac{R}{32 \pi  (2 C-1) \left(6 C+\ell ^2+\ell -2\right)^2}
    \bigg[-288 C^4-36 C^3 \left(\ell ^2+\ell -8\right)+12 C^2 (\ell -1) (\ell +2) \left(\ell ^2+\ell +3\right)\nonumber\\
    &+3 C \ell  (\ell +1) \left(\ell ^2+\ell -2\right)^2-\ell  (\ell +1) \left(\ell ^2+\ell -2\right)^2\bigg]
    \,,\\
    \mathscr{s}_2 &= 
    \frac{R^2 \left(24 C^2+6 C \left(\ell ^2+\ell -2\right)+(\ell -1) \ell  (\ell +1) (\ell +2)\right)}{32 \pi  \left(6 C+\ell ^2+\ell -2\right)}
    \,,\\
    \mathscr{s}_3 &= 
    \frac{R^2 (R-2 M)}{16 \pi }
    \,.
\end{align}
\end{subequations}
\end{widetext}

\section{Comparison to other definitions of overlap integrals in literature}\label{appendix:compare-overlap}
Here, we present a brief comparison of Eq.~\eqref{eq:Is-def} to the overlap integral prescriptions from~\cite{Kuan_2021,Miao_2024}.
The relativistic overlap prescriptions in these references are obtained through analogy to the Newtonian overlap prescription. For example, in Eq. (28) of~\cite{Kuan_2021}, the Newtonian ``tidal potential'' of the companion is used to compute their overlap integral. Such an object is well defined in Newtonian gravity, but not in general relativity. Moreover, Eq. (28) of~\cite{Kuan_2021} does not include the full self-gravity of the perturbed star in the overlap integral. 

If one makes the assumption of ignoring the self gravity of the star and using the Newtonian tidal potential, then Eq.~\eqref{eq:Is-def} reduces to Eq. (28) of~\cite{Kuan_2021}.
From Eq.~\eqref{eq:Is-def} we have
\begin{align}
    &\frac{4 \pi}{2 \ell + 1}d_{\ell m} I_{p}  
    \equiv 
    \frac{1}{M R^{{\ell}}}\int r^2 e^{\lambda/2} dr d \Omega
    e^{\nu(r)/2}
    \bigg[
    \xi_{\beta,p} \boldsymbol{F}^{\beta}[\boldsymbol{y}_{T}] 
    \nonumber\\
    &+ \frac{h_{\alpha \beta,p} \boldsymbol{F}^{\alpha \beta}[\boldsymbol{y}_{T}] }{16 \pi} 
    \bigg] 
    \label{eq:response_overlap_full}
    \,.
\end{align}
Here $\boldsymbol{F}_\beta$ and $\boldsymbol{F}_{\alpha\beta}$ are the effective fluid and
metric sources generated by the chosen tidal field $y_T$. The second term in the equation above is the part of the driving that comes from
the self-gravity of the perturbation. This term is absent in Eq. (28) of~\cite{Kuan_2021}.
To obtain their equation, first, one neglects the self-gravity contribution to the overlap
\begin{equation}
    h^{\alpha\beta}_p \boldsymbol{F}_{\alpha\beta}[y_T] \rightarrow 0 .
\end{equation}
Second, one identifies the applied tidal field with a Newtonian scalar
potential,
\begin{equation}
    U_T \propto 
    d_{\ell m} r^\ell Y_{\ell m} 
\end{equation}
In this limit, the effective force density entering the fluid equation becomes
\begin{equation}
    \boldsymbol{F}_\beta[y_T]
    \propto 
    (e+p)\nabla_\beta U_T
    \propto 
    d_{\ell m}(e+p)
    \nabla_\beta(r^\ell Y_{\ell m}) \,.
\end{equation}
Substituting this into Eq.~\eqref{eq:response_overlap_full} gives
\begin{align}
    &\frac{4\pi}{2\ell+1} d_{\ell m} I_p
    \rightarrow
    \nonumber\\
    &
    \frac{d_{\ell m}}{M R^\ell}
    \int r^2 e^{(\lambda+\nu)/2}drd\Omega\,
    (e+p)\xi^\beta_p
    \nabla_\beta(r^\ell Y_{\ell m}) .
\end{align}
For $\ell=m=2$ and the mode-index $p$ equal to the $f$-mode, this is exactly proportional to Eq.~(28) of~\cite{Kuan_2021}. We find that a similar argument results in Eq. (A15) of~\cite{Miao_2024}.
\bibliography{ref}
\end{document}